\documentclass[10pt]{article}  %envcountsame
\usepackage{amsmath}
\usepackage{amsfonts,amssymb}

\begin{document}

\newtheorem{theorem}{Theorem}[section]
\newtheorem{lemma}{Lemma}[section]
\newcommand{\bproof}{\smallskip{\em Proof. }}
\newcommand{\eproof}{\smallskip}
\newcommand{\qed}{{\hbox{$\  \square$}}}

\newcommand{\nl}{\nonumber\\}
\newcommand{\nnl}{\nl[6mm]}
\newcommand{\nle}{\nl[-2.0mm]\\[-2.0mm]}
\newcommand{\nlb}[1]{\nl[-2.0mm]\label{#1}\\[-2.0mm]}
\newcommand{\bl}{&&\quad}
\newcommand{\ab}{\allowbreak}

\renewcommand{\leq}{\leqslant}
\renewcommand{\geq}{\geqslant}

\renewcommand{\theequation}{\thesection.\arabic{equation}}
\let\ssection=\section
\renewcommand{\section}{\setcounter{equation}{0}\ssection}

\newcommand{\be}{\bes}
\newcommand{\ee}{\ees}
\newcommand{\bes}{\begin{eqnarray}}
\newcommand{\ees}{\end{eqnarray}}
\newcommand{\eens}{\nonumber\end{eqnarray}}

\renewcommand{\/}{\over}
\renewcommand{\d}{\partial}
\newcommand{\ddt}{{d\/dt}}

\newcommand{\eps}{\epsilon}
\newcommand{\dlt}{\delta}
\newcommand{\al}{\alpha}
\newcommand{\si}{\sigma}
\newcommand{\la}{\lambda}
\newcommand{\ka}{\kappa}
\newcommand{\emp}{\emptyset}

\newcommand{\xmu}{\xi^\mu}
\newcommand{\xnu}{\xi^\nu}
\newcommand{\xz}{\xi^0}
\newcommand{\ynu}{\eta^\nu}
\newcommand{\yz}{\eta^0}
\newcommand{\dmu}{\d_\mu}
\newcommand{\dnu}{\d_\nu}
\newcommand{\dsi}{\d_\si}
\newcommand{\dtau}{\d_\tau}
\newcommand{\drho}{\d_\rho}
\newcommand{\qmu}{q^\mu}
\newcommand{\qnu}{q^\nu}
\newcommand{\qsi}{q^\si}
\newcommand{\qtau}{q^\tau}
\newcommand{\qrho}{q^\rho}
\newcommand{\pmu}{p_\mu}
\newcommand{\pnu}{p_\nu}
\newcommand{\prho}{p_\rho}

\newcommand{\summ}{\sum_{\mm}}
\newcommand{\sumn}{\sum_{\nn}}
\newcommand{\sumr}{\sum_{\rr}}
\newcommand{\sums}{\sum_{\ss}}
\newcommand{\summs}{\sum_{\mm,\ss}}
\newcommand{\sumns}{\sum_{\nn,\ss}}
\newcommand{\sumrs}{\sum_{\rr,\ss}}
\newcommand{\summn}{\sum_{\mm,\nn}}
\newcommand{\summr}{\sum_{\mm,\rr}}
\newcommand{\sumnr}{\sum_{\nn,\rr}}
\newcommand{\summnr}{\sum_{\mm,\nn,\rr}}
\newcommand{\summns}{\sum_{\mm,\nn,\ss}}
\newcommand{\summrs}{\sum_{\mm,\rr,\ss}}
\newcommand{\summnrs}{\sum_{\mm,\nn,\rr,\ss}}

\newcommand{\dsum}[2]{\sum_{\scriptstyle{#1}\atop\scriptstyle{#2}}}

\newcommand{\summp}[1]{\sum_{|\mm|\leq p#1}}
\newcommand{\sumnp}[1]{\sum_{|\nn|\leq p#1}}
\newcommand{\sumrp}[1]{\sum_{|\rr|\leq p#1}}
\newcommand{\sumsp}[1]{\sum_{|\ss|\leq p#1}}
\newcommand{\summnp}[1]{\sum_{|\mm|,|\nn|\leq p#1}}
\newcommand{\sumnsp}[1]{\sum_{|\nn|,|\ss|\leq p#1}}

\newcommand{\xxy}{\xi\leftrightarrow\eta}
\newcommand{\fxg}{f\leftrightarrow g}

\newcommand{\phim}{\phi_{,\mm}}
\newcommand{\phin}{\phi_{,\nn}}
\newcommand{\pim}{{\pi}^{,\mm}}
\newcommand{\pin}{{\pi}^{,\nn}}

\newcommand{\tpi}{{1\/2\pi i}}
\newcommand{\ktpi}{{k^{\emp\emp}\/2\pi i}}
\newcommand{\fpi}{{1\/4\pi i}}
\newcommand{\half}{{1\/2}}
\newcommand{\Np}[1]{{N+p\choose N #1}}
\newcommand{\Npr}{{N+p-r\choose N-r}}

%
% Typesetter: I really mean mathbf here; confusion with 3-dimensional
% vectors is not possible.
%
\newcommand{\mm}{{\mathbf m}}
\newcommand{\nn}{{\mathbf n}}
\newcommand{\rr}{{\mathbf r}}
\renewcommand{\ss}{{\mathbf s}}
\newcommand{\uu}{{\mathbf u}}
\newcommand{\vv}{{\mathbf v}}

\newcommand{\one}{{\hat 1}}
\newcommand{\two}{{\hat 2}}

\renewcommand{\L}{{\cal L}}
\newcommand{\J}{{\cal J}}
\newcommand{\UU}{{\cal U}}
\newcommand{\EE}{{\cal E}}
\newcommand{\FF}{{\cal F}}
\newcommand{\Lxi}{\L_\xi}
\newcommand{\Leta}{\L_\eta}

\newcommand{\into}{\hookrightarrow}
\newcommand{\e}{{\rm e}}
\newcommand{\ext}{{\rm ext}}
\newcommand{\larroww}[1]{{\ \stackrel{#1}{\longleftarrow}\ }}
\renewcommand{\th}{^{{\rm th}}}

\newcommand{\kzero}{k^{(0)}}
\newcommand{\kone}{k^{(1)}}
\newcommand{\ki}{k^{(i)}}
\newcommand{\kj}{k^{(j)}}
\newcommand{\kl}{k^{(\ell)}}
\newcommand{\kr}{k^{(r)}}

\newcommand{\dzero}{d^{(0)}}
\newcommand{\done}{d^{(1)}}
\newcommand{\di}{d^{(i)}}

\newcommand{\rep}{{\varrho}}

\newcommand{\mri}{(-)^i {r\choose i}}

\newcommand{\tr}{{\rm tr}\kern0.7mm}
\newcommand{\oj}{{\mathfrak g}}
					    
\renewcommand{\gg}{\oj\oplus gl(N)}
\newcommand{\Mg}{{\UU_M}(\oj,N,p)}
\newcommand{\Mgg}{{\UU_M}(\gg,N,p)}
\newcommand{\Ug}{\UU(\oj,N,p)}
\newcommand{\Ugg}{\UU(\gg,N,p)}

\newcommand{\Lg}{\widehat{\oj}}
\newcommand{\Lgl}{\widehat{gl}}
\newcommand{\LUU}{\widehat\UU}
\newcommand{\LMg}{\widehat{\UU_M}(\oj,N,p)}
\newcommand{\LMgg}{\widehat{\UU_M}(\gg,N,p)}
\newcommand{\LUg}{\widehat\UU(\oj,N,p)}
\newcommand{\LUgg}{\widehat\UU(\gg,N,p)}

\newcommand{\bra}[1]{\big{\langle}#1\big{|}}
\newcommand{\ket}[1]{\big{|}#1\big{\rangle}}

\newcommand{\no}[1]{{\,:\kern-0.7mm #1\kern-1.2mm:\,}}

\newcommand{\RR}{{\mathbb R}}
\newcommand{\CC}{{\mathbb C}}
\newcommand{\ZZ}{{\mathbb Z}}
\newcommand{\NN}{{\mathbb N}}

\title{{Multi-dimensional Diffeomorphism and Current Algebras from
Virasoro and Kac-Moody Currents}}
\author{T. A. Larsson \\
Vanadisv\"agen 29, S-113 23 Stockholm, Sweden\\
email: thomas.larsson@hdd.se}

\maketitle
\begin{abstract}
The recently constructed Fock representations of $N$-dimensional 
diffeomorphism and current algebras are reformulated in terms of
one-dimensional currents, satisfying Virasoro and affine Kac-Moody
algebras. 
\end{abstract}

\section{Introduction}

In a recent paper \cite{Lar98}, I constructed Fock representations of 
diffeomorphism and current algebras in $N$-dimensional spacetime. 
More precisely, I considered the
{\em DGRO (Diffeomorphism, Gauge, Repara\-metrization, Observer)} algebra
\break $DGRO(N,\oj)$, where $\oj$ is a finite-dimensional Lie algebra.
The Fock representations were obtained by expanding functions, valued
in suitable spaces, in a multi-dimensional Taylor series around the
points of a one-dimensional trajectory, prior to normal ordering.
In the resulting expressions, the Fock oscillators only appear in
bilinear combinations. Therefore, it suffices to find currents that
satisfy the same algebraic relations as these bilinears.
In the present paper I write down the relevant expressions in terms of
affine Kac-Moody and Virasoro currents, and verify that they indeed 
satisfy $DGRO(N,\oj)$. This means that new representions of the full DGRO
algebra can be constructed using representations of these Kac-Moody and 
Virasoro algebras. In particular, I give a Sugawara construction for
the reparametrization subalgebra.

The construction works for every finite jet order $p$ (the order at which
the Taylor expansion is truncated),
but the abelian charges, i.e. the parameters multiplying the cocycles,
diverge when $p\to\infty$; the worst parameters behave like $p^{N+2}$.
An important problem is to construct factor modules such that
the limit $p\to\infty$ exists, at least for $N$ sufficiently small and
for some choice of $\oj$. A first step in this direction is taken in 
Section \ref{sec:finite}. In the direct sum of realizations with jet 
order ranging from $p-r$ to $p$, the leading divergences can be made to 
cancel. Then the abelian charges are independent of $p$ for $N=r$ and 
they vanish for $N<r$.

The higher-dimensional analogs of Virasoro and loop algebras
have not attracted much attention compared to their one-dimensional
siblings. One reason may be that Pressley and Segal, 
in their influential book {\em Loop groups} \cite{PS86},
noted that the higher-dimensional Kac-Moody-like cocycle follows by 
pull-back from the one-dimensional case. Equivalently, the extension
of $map(N,\oj)$, which is the algebra of $\oj$-valued functions, is 
obtained by restriction of these functions to some priviledged 
one-dimensional curve ("the observer's trajectory"). 
However, I believe that their claim is somewhat misleading,
because it does not imply that all modules are inherited
from $\Lg$. 
The realizations constructed in the present 
paper depend on more data: not only do they involve functions on the 
preferred curve, but also derivatives of these functions up to some 
fixed finite order $p$. Since this includes transverse derivatives, 
it is a truly higher-dimensional effect. 
Moreover, these realizations are expressed in
terms of Kac-Moody currents, but the relevant algebra is not
$\widehat{\oj}$, except as a special case.

For diffeomorphisms the difference between $N=1$ and $N>1$ dimensions
is even more dramatic. The observer's trajectory is not preserved by
diffeomorphisms, and thus the higher-dimensional Virasoro extensions
are not central; in fact, the current algebra cocycle does not commute 
with diffeomorphisms neither. Note also that the classical $diff(N)$
modules (tensor densities) depend crucially on the dimension.

Related work can be found in 
\cite{BB98,Bil97,Lar97,Lar98,Lar99,MRY90,RMY92,RM94}.
Cocycles of the diffeomorphism algebra were classified by
Dzumadil'daev \cite{Dzhu96} and reviewed in \cite{Lar00}.

\section{Background}

\subsection{DGRO Algebra}

Let $\xi=\xmu(x)\dmu$, $x\in\RR^N$, $\dmu = \d/\d x^\mu$,
be a vector field, with commutator 
$[\xi,\eta] \equiv \xmu\dmu\ynu\dnu - \ynu\dnu\xmu\dmu$.
Greek indices $\mu,\nu = 1,2,..,N$ label the 
spacetime coordinates and the summation convention is used on all kinds 
of indices.
The diffeomorphism algebra (algebra of vector fields, Witt algebra) 
$diff(N)$ is generated by Lie derivatives $\Lxi$.
In particular, we refer to diffeomorphisms on the circle as 
repara\-metrizations. They form an additional $diff(1)$ algebra with 
generators $L_f$, where $f = f(t)d/dt$, $t\in S^1$, is a vector field 
on the circle.

Let $\oj$ be a finite-dimen\-sional Lie algebra with basis $J^a$
(hermitian if $\oj$ is compact and semisimple), structure constants 
$f^{ab}{}_c$, and Killing metric $\dlt^{ab}$. The brackets are
\be
[J^a,J^b] = if^{ab}{}_c J^c.
\label{g}
\ee
Let $\dlt^a\propto\tr J^a$ be proportional to the linear Casimir 
operator, satisfying $f^{ab}{}_c\dlt^c \equiv 0$. Clearly,
$\dlt^a=0$ if $\oj$ is semisimple, but it may be non-zero on abelian 
factors. Let $map(N,\oj)$ denote the {\em gauge algebra} of maps from
$N$-dimen\-sional spacetime to $\oj$. It is the algebra of
$\oj$-valued functions $X = X_a(x)J^a$ with commutator
$[X,Y] \equiv if^{ab}{}_c X_a Y_b J^c$. Denote its generators by $\J_X$.
The action of $diff(N)$ on $map(N,\oj)$ is given by
$\xi X =\xmu\dmu X_aJ^a$.

Finally, let the $Obs(N)$ be the space of local functionals of the
observer's tractory $\qmu(t)$, i.e. polynomial functions of  
$\qmu(t)$, $\dot\qmu(t)$, ... $d^k \qmu(t)/dt^k$,  $k$ finite, 
regarded as a commutative algebra. $Obs(N)$ is a $diff(N)$ module in a 
natural manner.

The DGRO algebra $DGRO(N,\oj)$ is an abelian but non-central Lie algebra 
extension of $diff(N) \ltimes map(N,\oj) \oplus diff(1)$ by $Obs(N)$:
\[
0 \longrightarrow Obs(N) \longrightarrow DGRO(N,\oj) \longrightarrow
 diff(N)\ltimes map(N,\oj)\oplus diff(1) \longrightarrow 0.
\]
The extension depends on the eight parameters $c_j$, $j = 1, ..., 8$,
to be called  {\em abelian charges}. The brackets are given by
\bes
[\Lxi,\Leta] &=& \L_{[\xi,\eta]} 
 + \tpi\int dt\ \dot\qrho(t) 
 \Big\{ c_1 \drho\dnu\xmu(q(t))\dmu\ynu(q(t)) +\nl
 \bl+ c_2 \drho\dmu\xmu(q(t))\dnu\ynu(q(t)) \Big\}, \nl
{[}\Lxi, \J_X] &=& \J_{\xi X} 
- {c_7 \/2\pi i} \dlt^a\int dt\ \dot\qrho(t) X_a(q(t))\drho\dmu\xmu(q(t)), \nl
{[}\J_X, \J_Y] &=& \J_{[X,Y]} - \tpi(c_5\dlt^{ab}+c_8\dlt^a\dlt^b) 
 \int dt\ \dot\qrho(t)\drho X_a(q(t))Y_b(q(t)), \nl
{[}L_f, \Lxi] &=& {c_3\/4\pi i} \int dt\ 
 (\ddot f(t) - i\dot f(t))\dmu\xmu(q(t)), 
\label{DGRO} \\
{[}L_f,\J_X] &=& {c_6 \/4\pi i}\dlt^a 
 \int dt\ (\ddot f(t) - i \dot f(t)) X_a(q(t)), \nl
{[}L_f,L_g] &=& L_{[f,g]} 
 + {c_4\/24\pi i}\int dt (\ddot f(t) \dot g(t) - \dot f(t) g(t)), \nl
{[}\Lxi,\qmu(t)] &=& \xmu(q(t)), \nl
{[}L_f,\qmu(t)] &=& -f(t)\dot\qmu(t), \nl
{[}\J_X, \qmu(t)] &=& {[}\qmu(s), \qnu(t)] = 0,
\eens
extended to all of $Obs(N)$ by Leibniz' rule and linearity.

\subsection{ Tensor fields and Fock modules }
In \cite{Lar98} Fock representations of $DGRO(N,\oj)$ were constructed. 
We started from from classical fields transforming as
\bes
[\Lxi, \phi(x,t)] &=& -\xmu(x)\dmu\phi(x,t)
- \dnu\xmu(x,t)T^\nu_\mu\phi(x,t), \nl
{[}\J_X, \phi(x,t)] &=& -X_a(x)J^a\phi(x,t), 
\label{tensor}\\
{[}L_f, \phi(x,t)] &=& -f(t)\dot\phi(x,t) 
- \la\dot f(t)\phi(x,t) + iw f\phi(x,t).
\eens
Here $J^a$ satisfies $\oj$ (\ref{g}), and $T^\mu_\nu$ satisfies
$gl(N)$, with brackets
\be
[T^\mu_\nu, T^\rho_\si] = 
 \dlt^\rho_\nu T^\mu_\si - \dlt^\mu_\si T^\rho_\nu,
\label{glN}
\ee
and $\phi(x,t)$ takes values in some $\gg$ module.
Let $\mm = (m_1, \ab m_2, \ab ..., \ab m_N)$, all $m_\mu\geq0$, be a 
multi-index of length $|\mm| = \sum_{\mu=1}^N m_\mu$. 
Denote by $\mu$ a unit vector in the $\mu\th$ direction, so that
$\mm+\mu = (m_1, \ab ...,m_\mu+1, \ab ..., \ab m_N)$, and let
\be
\phim(t) = \d_\mm\phi(q(t),t)
= \underbrace{\d_1 .. \d_1}_{m_1} .. 
\underbrace{\d_N .. \d_N}_{m_N} \phi(q(t),t)
\label{jetdef}
\ee
be the $|\mm|\th$ order derivative of $\phi(x,t)$ on the
observer's trajectory $\qmu(t)$. Such objects transform as
\bes
[\Lxi, \phim(t)] &=& \d_\mm([\Lxi,\phi(q(t),t)]) 
+ [\Lxi,\qmu(t)]\dmu\d_\mm\phi(q(t),t) \nl
&=& -\sumn T^\nn_\mm(\xi(q(t))) \phin(t), \nl
{[}\J_X, \phim(t)] &=& \d_\mm([\J_X,\phi(q(t),t)]) 
\label{jet} \\
&=& -\sumn J^\nn_\mm(X(q(t))) \phin(t), \nl
{[}L_f, \phim(t)] &=& -f(t)\dot\phim(t) 
- \la\dot f(t)\phim(t) + iw f\phim(t),
\eens
where
\bes
T^\mm_\nn(\xi) &=& 
 {\nn\choose\mm} \d_{\nn-\mm+\nu}\xmu T^\nu_\mu \nl
 &+& {\nn\choose\mm-\mu}\d_{\nn-\mm+\mu}\xmu
  - \dlt^\mm_{\nn+\mu} \xmu,
\label{Tmn}\\
J^\mm_\nn(X) &=& {\nn\choose\mm} \d_{\nn-\mm} X_a J^a.
\eens
Here $\mm! = m_1!m_2!...m_N!$ and
\be
{\mm\choose\nn} = {\mm!\/\nn!(\mm-\nn)!} = 
{m_1\choose n_1}{m_2\choose n_2}...{m_N\choose n_N}.
\ee
Some other properties of multi-dimensional binomial coefficients
are listed in appendix \ref{sec:A}.

Here and henceforth we use the convention that a sum over
a multi-index runs over all values of length at most $p$. Since
$T^\nn_\mm(\xi)$ and $J^\nn_\mm(X)$ vanish whenever $|\nn|>|\mm|$,
the sums over $\nn$ in (\ref{jet}) are in fact further restricted.

Add dual coordinates (jet momenta) $(\pmu(t), \pim(t))$, which
satisfy
\bes
[\pmu(s), \qnu(t)] &=& \dlt^\nu_\mu \dlt(s-t), 
\label{Poisson} \\
{\{}\pim(s), \phin(t)\} &\equiv& - \{\phin(t), \pim(s)\}
 = \dlt^\mm_\nn \dlt(s-t),
\eens
and all other brackets vanish. For definiteness, we take the fields
to be fermionic, as indicated by the curly brackets.
Then it follows immediately from (\ref{jet}) that the following
operators define a realization of $DGRO(N,\oj)$ in Fock space:
\bes
\Lxi &=& \int dt\ \Big\{ \no{\xmu(q(t)) \pmu(t)} +
\summn \no{ \pim(t) T^\nn_\mm(\xi(q(t))) \phin(t)  } \Big\}, \nl
\J_X &=& \int dt\ \summn \no{ \pim(t) J^\nn_\mm(X(q(t))) \phin(t) }, 
\label{Fock} \\
L_f &=& \int dt\ \Big\{ - f(t) \no{\dot\qmu(t)\pmu(t)} +\nl
&&+ \summ \no{\pim(t)(f(t)\dot\phim(t) 
+ \la\dot f(t)\phim(t) - iw f\phim)} \Big\},
\eens
where double dots ($\no{\quad}$) denote normal ordering with respect to 
frequency.

\section{ Virasoro / Kac-Moody Realization }
By means of (\ref{Tmn}), we can rewrite (\ref{Fock}) as
\bes
\Lxi &=& \int dt\ \Big\{ \no{\xmu(q(t)) \pmu(t)} 
 - \xmu(q(t)) P_\mu(t) +\nl
 &&+\summn {\mm\choose \nn}
  \d_{\mm-\nn}\xmu(q(t)) E^\mm_{\nn+\mu}(t) \Big\}
 + T_{d\xi}, \nl
T_{d\xi} &=& \int dt\ \summn {\mm\choose \nn} 
 \d_{\mm-\nn+\nu}\xmu(q(t)) T^{\mm\nu}_{\nn\mu}(t), 
\label{DROreal} \\
J_X  &=& \int dt\ \summn {\mm\choose \nn} 
 \d_{\mm-\nn}X_a(q(t)) J^{\mm a}_\nn(t), \nl
L_f &=& \int dt\ f(t)L(t),
\eens
where 
\bes
P_\mu(t) = \summ E^\mm_{\mm+\mu}(t), \nlb{PL}
L(t) = - \no{\dot\qmu(t)\pmu(t)} + F(t),
\eens
and
\bes
E^\mm_\nn(t) &=& \no{ \pim(t)\phin(t) }, \nl
J^{\mm a}_\nn(t) &=& \no{ \pim(t) J^a \phin(t) }, 
\nlb{Freal}
T^{\mm\nu}_{\nn\mu}(t) &=& \no{ \pim(t) T^\mu_\nu \phin(t) }, \nl
F(t) &=& \summ \no{\pim(t)\dot\phim(t)}.
\eens

The currents $E^\mm_\nn(t)$ satisfy the Kac-Moody algebra 
$\Lgl(\Np{})$,\footnote{To avoid unduly wide hats, I put the Kac-Moody 
hat only over the algebra's given name.} with brackets
\be
[E^\mm_\nn(s), E^\rr_\ss(t)] =
(\dlt^\rr_\nn E^\mm_\ss(s) - \dlt^\mm_\ss E^\rr_\nn(s)) \dlt(s-t)
+ \ktpi \dlt^\mm_\ss\dlt^\rr_\nn \dot\dlt(s-t).
\ee
The notation $k^{\emp\emp}$ for the central charge 
will be explained shortly. In particular,
\bes
[P_\mu(s), P_\nu(t)] &=& 0, \nle
{[}P_\mu(s), E^\mm_\nn(t)] &=& 
 (E^{\mm-\mu}_\nn(s) - E^\mm_{\nn+\mu}(s))\dlt(s-t)
+ \ktpi \dlt^\mm_{\nn+\mu} \dot\dlt(s-t), 
\eens

However, the currents $J^{\mm a}_\nn(t)$ and $T^{\mm\nu}_{\nn\mu}(t)$
do not satisfy such a simple algebra. They span a vector
space which is isomorphic to ${\oj\oplus gl(\Np{})}$
and ${gl(N)\oplus gl(\Np{})}$, respectively, but these
spaces are not preserved by the bracket. Indeed,
\be
[J^{\mm a}_\nn, J^{\rr b}_\ss] &=&
\half \dlt^\rr_\nn (if^{ab}{}_c J^{\mm a}_\ss + I^{\mm (ab)}_\ss)
- \half \dlt^\mm_\ss(if^{ab}{}_c J^{\rr a}_\nn + I^{\rr (ab)}_\nn),
\ee
where $I^{\mm (ab)}_\nn = \pim (J^aJ^b + J^bJ^a) \phin$ contains
the symmetric expression $J^aJ^b + J^bJ^a$, belonging to $\UU(\oj)$,
the universal enveloping algebra of $\oj$.
$\UU(\oj)$ has basis $I^A$, where $A = (), (a), (a_1a_2), \ldots,
(a_1a_2\ldots a_n), \ldots$ consists of $n$-tuples of $\oj$ indices.
These tuples are completely symmetric; any anti-symmetry can always
be expressed in terms of lower-order tuples by means of (\ref{g}).
A typical element in $\UU(\oj)$ thus has the form
$I^{(a_1a_2\ldots a_n)} = J^{(a_1} J^{a_2} \ldots J^{a_n)}$,
where parentheses denote symmetrization.
We identify $\oj$ with the one-tuples in $\UU(\oj)$, $J^a = I^{(a)}$,
and denote the empty set by $()=\emp$. By construction, $\UU(\oj)$ 
is an associative
algebra, and we denote its structure constants by $g^{AB}{}_C$:
\be
I^A I^B = g^{AB}{}_C I^C.
\label{Ug}
\ee
E.g., $g^{(a)(b)}{}_{(c)} = -g^{(b)(a)}{}_{(c)} = \half if^{ab}{}_c$,  
$g^{(a)(b)}{}_{(cd)} = \half\dlt^{(ab)}{}_{(cd)}$,
$g^{ab}{}_C = 0$ otherwise, and
$g^{A\emp}{}_B = g^{\emp A}{}_B = \dlt^A_B$.
For brevity, we shall not parenthesize single $\oj$ indices henceforth.

Define $\Ug$ as the Lie algebra with basis $I^{\mm A}_\nn$,
where $A$ is a $\UU(\oj)$ index and $\mm$, $\nn$ are multi-indices
of length at most $p$. The brackets read
\be
[I^{\mm A}_\nn, I^{\rr B}_\ss] =
\dlt^\rr_\nn g^{AB}{}_C I^{\mm C}_\ss
- \dlt^\mm_\ss g^{BA}{}_C I^{\rr C}_\nn.
\label{Ugp}
\ee
That this is a Lie algebra follows from the explicit realization
$I^{\mm A}_\nn = \pim I^A \phin$.
Note that $\Ug$ is a highly reducible Lie algebra. It contains $\Np{}$
different subalgebras isomorphic to $\oj$, generated by $I^{\mm(a)}_\mm$
(no sum on $\mm$), as well as a $gl(\Np{})$ subalgebra generated by
$I^{\mm \emp}_\nn$.

$\UU(\oj)$ has the universal property that every $\oj$ representation
$M$ is given by a homomorphism $\UU(\oj) \to M$. We can therefore 
reinterpret $A,B$ as $M$ indices, view $I^A$ as a basis for $M$, and let
$g^{AB}{}_C$ be the structure constants for the associative product
in $M$. The brackets (\ref{Ugp}) still define a Lie algebra, which
we denote by $\Mg$. Contrary to $\Ug$, this algebra
is finite-dimensional provided that $M$ is a finite-dimensional
representation of $\oj$.

The corresponding Kac-Moody algebras are $\LUg$ and $\LMg$. A basis is
given by $I^{\mm A}_\nn(t)$, $t\in S^1$, and the brackets read
\bes
[I^{\mm A}_\nn(s), I^{\rr B}_\ss(t)] &=&
\dlt^\rr_\nn g^{AB}{}_C I^{\mm C}_\ss(s) \dlt(s-t)
- \dlt^\mm_\ss g^{BA}{}_C I^{\rr C}_\nn(s) \dlt(s-t) +\nl
&&+ {k^{AB}\/2\pi i} \dlt^\mm_\nn \dot\dlt(s-t).
\label{KMJ}
\ees
If $\Ug$ were semi-simple, the {\em central charge matrix} $k^{AB}$ 
would be proportional to $\dlt^{AB}$,
but this needs not be true in the presence of a linear Casimir.

Of particular interest is the case that $\oj=gl(N)$, due to the
appearence of $gl(N)$ representations in $diff(N)$ representations
(tensor fields). This is obtained from the general case by substituting
$J^a \mapsto T^\mu_\nu$, so a $gl(N)$ basis is labelled by duplets
${}^\mu_\nu$. 
More generally, we need the direct sum $\gg$, with basis $J^a$ and 
$T^\mu_\nu$. The $\gg$ basis is thus labelled by ${}^{\mu a}_\nu$.
Now reinterpret the $\UU(\oj)$ labels $A$ as $\UU(\gg)$ labels, so
a typical element in $\Ugg$ has the form 
$I^{\mm A}_\nn = I^{\mm(\mu_1..\mu_k)(a_1...a_\ell)}_{\nn(\nu_1..\nu_k)}$.
It is labelled by multi-indices $\mm$, $\nn$, symmetric tuples of $\oj$ 
indices $(a_1...a_\ell)$ and tuples of $gl(N)$ indices
${}^{(\mu_1..\mu_k)}_{(\nu_1..\nu_k)}$, symmetric under simultaneuous
interchange of pairs $(\mu_i,\nu_i) \leftrightarrow (\mu_j,\nu_j)$.
The three first operators in (\ref{Freal}) are embedded into the
corresponding Kac-Moody algebra (\ref{KMJ}) as follows:
\bes
E^\mm_\nn &=& I^{\mm()()}_{\nn()} \equiv I^{\mm\emp}_\nn, \nl
J^{\mm a}_\nn &=& I^{\mm()(a)}_{\nn()}, 
\label{EJT}\\
T^{\mm\mu}_{\nn\nu} &=& I^{\mm(\mu)()}_{\nn(\nu)}.
\eens

To reduce writing, we introduce an abbreviated notation where a label
$M = {}^{\mm A}_\nn$ (a latin capital from the middle of the alphabeth)
stands for both $gl(\Np{})$ multi-indices and $\UU(\gg)$
indices. The bracket in $\Ugg$ can now be written as
\be
[I^M, I^N] = if^{MN}{}_R I^R,
\ee
and the corresponding Kac-Moody algebra $\LUgg$ reads
\be
[I^M(s), I^N(t)] = if^{MN}{}_R I^R(s)\dlt(s-t)
 + { k^{MN}\/2\pi i} \dot\dlt(s-t).
\label{KMI}
\ee
Skew-symmetry and the Jacobi identities imply the relations
\bes
k^{NM} &=& k^{MN}, 
\nlb{cond}
f^{MN}{}_S k^{SR} &=& f^{NR}{}_S k^{SM} = f^{RM}{}_S k^{SN}.
\eens
The structure constants are then given by
\bes
if{}^{\mm A}_\nn{}^{\rr B}_\ss{}_{\uu C}^\vv
&=& \dlt^\rr_\nn \dlt^\mm_\uu \dlt^\vv_\ss g^{AB}{}_C
- \dlt^\mm_\ss \dlt^\rr_\uu \dlt^\vv_\nn g^{BA}{}_C, \nle
k{}^{\mm A}_\nn{}^{\rr B}_\ss 
&=& \dlt^\rr_\nn \dlt^\mm_\ss k^{AB}.
\eens
The conditions (\ref{cond}) take the form
\bes
k^{AB} &=& k^{BA}, \nle
g^{AB}{}_D k^{DC} &=& g^{BC}{}_D k^{DA} = g^{CA}{}_D k^{DB}.
\eens
We are mainly interested in what restrictions these conditions impose
on the first few elements of the central charge matrix. For the sake of
this argument, we temporarily ignore the $gl(N)$ factor and only consider 
$\oj$. Specialization 
to $A=(a)$, $B=C=\emp$ gives $k^{\emp a} = k^{a\emp}$,
which is already clear. The case $A=(a)$, $B=(b)$, $C=\emp$ is more
interesting:
\be
k^{ab} = k^{ba} = g^{ab}{}_\emp k^{\emp\emp}
 + g^{ab}{}_c k^{c\emp} + g^{ab}{}_{(cd)} k^{(cd)\emp}.
\ee
The symmetric part gives a condition on $k^{(ab)\emp}$:
\be
k^{(ab)\emp} = k^{ab} + k^{ba}
 - (g^{ab}{}_\emp +g^{ba}{}_\emp)k^{\emp\emp},
\ee
which we do not need here. The skew part implies that
$f^{ab}{}_c k^{c\emp} = 0$, i.e. that $k^{c\emp} \propto \delta^c$,
the linear Casimir.
Finally, the case $A=(a)$, $B=(b)$, $C=(c)$ gives
\be
g^{ab}{}_D k^{Dc} &=& g^{bc}{}_D k^{Da} = g^{ca}{}_D k^{Db}.
\label{gk}
\ee
The even part of this condition yields
\be
(g^{ab}_\emp+g^{ba}_\emp)k^{\emp c} + k^{(ab)c}
= (g^{bc}_\emp+g^{cb}_\emp)k^{\emp a} + k^{(bc)a},
\ee
which constrains $k^{(ab)c}$.
The odd part leads to the usual Kac-Moody condition 
$f^{bc}{}_d k^{da} = f^{ab}{}_d k^{dc}$. One solution is the usual
one proportional to the quadratic Casimir: $k^{ab} = k\dlt^{ab}$, but
there is also another solution due to the presence of a linear
Casimir: $k^{ab} \propto \dlt^a\dlt^b$.

We do not need the full central charge matrix to determine the abelian
charges of the DGRO algebra, but only the components generated from
the operators (\ref{EJT}). The conditions we just studied imply that
these components must be of the form
\bes
k^{\mu\rho}_{\nu\si} &=& k_1 \dlt^\mu_\si\dlt^\rho_\nu
 + k_2 \dlt^\mu_\nu\dlt^\rho_\si, \nl
k^{\mu\emp}_\nu &=& k_3 \dlt^\mu_\nu, \nl
k^{\emp\emp} &=& k_4, 
\nlb{ks}
k^{ab} &=& k_5 \dlt^{ab} + k_8 \dlt^a\dlt^b, \nl
k^{a\emp} &=& k_6\dlt^a, \nl
k^{\mu a}_\nu &=& k_7 \dlt^\mu_\nu \dlt^a.
\eens
Equation (\ref{ks}) defines eight independent parameters $k_1 - k_8$, 
which characterize
the Kac-Moody algebra $\LUgg$, at least in part. Its central charge 
matrix may have additional independent components, but if so they arise 
for higher-order operators, which do not affect the abelian
charges of the DGRO algebra.

The realization (\ref{DROreal}) also depends on a fourth operator $F(t)$,
which generates the Virasoro algebra $Vir$ commuting with the observer's
trajectory:
\bes
[F(s),F(t)] &=& (F(s)+F(t)) \dot\dlt(s-t) 
 + {c(N,p)\/24\pi i}(\dddot\dlt(s-t) + \dot\dlt(s-t)),
\label{Vir} \\
{[}F(s),\qmu(t)] &=& [F(s),\pmu(t)] = 0.
\label{commute}
\ees
The generators of $\LUgg$ transform as weight one primary fields:
\be
[F(s),I^M(t)] &=& I^M(s)\dot\dlt(s-t)
+ {d^M\/4\pi i}(\ddot\dlt(s-t)+i\dot\dlt(s-t)).
\label{LI}
\ee
The $LII$ Jacobi identity leads to the condition
$f^{MN}{}_R d^R = 0$, which means that the parameters $d^M$ are
proportional to the linear Casimir. In particular,
\bes
d^\emp &=& d_0, \nl
d^\mu_\nu &=& d_1 \dlt^\mu_\nu, 
\label{ds}\\
d^a &=& d_2 \dlt^a,
\eens
which defines three additional parameters $d_0 - d_2$.
In particular, for the operators in (\ref{EJT}), (\ref{LI}) takes the
form
\bes
{[}F(s), E^\mm_\nn(t)] &=& E^\mm_\nn(s) \dot\dlt(s-t)
 + {d_0\/4\pi i} \dlt^\mm_\nn (\ddot\dlt(s-t)+i\dot\dlt(s-t)), \nl
{[}F(s), J^{\mm a}_\nn(t)] &=& J^{\mm a}_\nn(s) \dot\dlt(s-t)
 + {d_2\/4\pi i} \dlt^a \dlt^\mm_\nn (\ddot\dlt(s-t)+i\dot\dlt(s-t)), 
\label{LEJT}\\
{[}F(s), T^{\mm \mu}_{\nn\nu}(t)] &=& 
T^{\mm \mu}_{\nn\nu}(s) \dot\dlt(s-t)
+ {d_1\/4\pi i} \dlt^\mu_\nu \dlt^\mm_\nn(\ddot\dlt(s-t)+i\dot\dlt(s-t)).
\eens

We have thus rewritten the Fock realization in terms of currents
(\ref{Freal}), satisfying (\ref{KMI}), (\ref{Vir}) and (\ref{LI}).
In this process, all explicit reference to the Fock operators
$\pim(t)$ and $\phin(t)$ has vanished. The result is summarized in the
following theorem.

\begin{theorem}
\label{T1}
Embed $E^\mm_\nn(t)$, $J^{\mm a}_\nn(t)$, $T^{\mm \mu}_{\nn\nu}(t)$
and $F(t)$ into $Vir \ltimes \ab \LUgg$, with brackets given by
(\ref{KMI}), (\ref{Vir}) and (\ref{LI}), by means for (\ref{EJT}).
Then the generators (\ref{DROreal}) satisfy the DGRO algebra
$DGRO(N,\oj)$ (\ref{DGRO}). The abelian charges are given by
\bes
c_1 &=& 1 - k_1\Np{} - k_4 {N+p+1\choose N+2}, \nl
c_2 &=& - k_2\Np{} - 2k_3 \Np{+1} - k_4 {N+p\choose N+2}, \nl
c_3 &=&	1 + d_1\Np{} + d_0\Np{+1}, \nl
c_4 &=& 2N+c(N,p), 
\label{cs}\\
c_5 &=& k_5\Np{}, \nl
c_6 &=& d_2\Np{}, \nl
c_7 &=& k_7\Np{} + k_6\Np{+1}, \nl
c_8 &=& k_8\Np{},
\eens
where the parameters $k_1-k_8$ are defined in (\ref{ks}) and $d_0-d_2$
are defined in (\ref{ds}).
\end{theorem}

\bproof
It is clear that we have some realization of $DGRO(N,\oj)$, because the
Fock operators only enter through normal-ordered bilinear combinations.
Appendix \ref{sec:Pf1} contains an independent verification of all 
representation conditions and a calculation of the abelian charges.
\qed
\eproof

An even more general realization is obtained by the redefinition
$F(t) \mapsto F(t) + \Delta F(t)$, where
\be
\Delta F(t) &=& -\la(\dot D(t) + i D(t)),
\label{DF}
\ee
where $\la$ is a parameter and $D(t) = \summ E^\mm_\mm(t)$.
Hence $\Delta L_f = \la \int dt\ (\dot f(t) - if(t)) D(t)$.
One checks that $D(t)$ obeys
\bes
{[}F(s), D(t)] &=& D(s) \dot\dlt(s-t)
 + {d_0\/4\pi i} \Np{} (\ddot\dlt(s-t)+i\dot\dlt(s-t)) \nle
{[}D(s), I^M(t)] &=& {k^{M\emp}\/2\pi i}\dot\dlt(s-t).
\eens
In particular,
\bes
[D(s),D(t)] &=& 
 {k_4\/2\pi i} \Np{} \dot\dlt(s-t), \nl
{[}D(s), P_\mu(t)] &=& 0, \nl
{[}D(s), E^\mm_\nn(t)] &=& 
 {k_4\/2\pi i} \dlt^\mm_\nn \dot\dlt(s-t), \\
{[}D(s), J^{\mm a}_\nn(t)] &=& 
 {k_6\/2\pi i} \dlt^a \dlt^\mm_\nn \dot\dlt(s-t), \nl
{[}D(s), T^{\mm\mu}_{\nn\nu}(t)] &=& 
 {k_3\/2\pi i} \dlt^\mu_\nu \dlt^\mm_\nn \dot\dlt(s-t).
\eens
This leads to a change of some abelian charges:
\bes
\Delta c_3 &=& -2\la (k_3\Np{} + k_4\Np{+1}), \nl
\Delta c_4 &=& 12(\la d_0-\la^2 k_4)\Np{}, 
\nlb{Dc}
\Delta c_6 &=& -2\la k_6\Np{},
\eens
which can be accomodated by a shift in the parameters $d_0-d_2$ and
$c(N,p)$:
\bes
\Delta d_0 = -2\la k_4 &\qquad&
\Delta d_2 = -2\la k_6 \nle
\Delta d_1 = -2\la k_3 &\qquad&
\Delta c(N,p) = 12(\la d_0-\la^2 k_4)\Np{}.
\eens

\section{ Comparison with previous work }
The results in \cite{Lar98} are recovered by choosing the fermionic Fock 
realization (\ref{Freal}) for the Kac-Moody and Virasoro generators.
Let $\rep$ be a $gl(N)$ representation and $M$ a $\oj$ representation.
In \cite{Lar98} I defined numbers $k_0(\rep)$, $k_1(\rep)$, $k_2(\rep)$, 
$y_M$, $z_M$ and $w_M$ by the following relations:
\bes
\tr\,1 &=& \dim\,\rep\,\dim\,M, \nl
\tr\,T^\mu_\nu &=& k_0(\rep)\, \dim\,M \dlt^\mu_\nu,\nl
\tr\,T^\mu_\nu T^\si_\tau &=& (k_1(\rep)\dlt^\mu_\tau\dlt^\si_\nu
+ k_2(\rep)\dlt^\mu_\nu\dlt^\si_\tau)\dim\,M, 
\nle
\tr\,J^a &=& z_M\dim\,\rep\,\dlt^a, \nl
\tr\,J^aJ^b &=& (y_M\dlt^{ab} + z_M\dlt^a\dlt^b)\dim\,\rep, \nl
\tr\,J^aT^\mu_\nu &=& z_Mk_0(\rep)\dlt^\mu_\nu\dlt^a,
\eens
where the trace is taken in the $\gg$ representation $M\oplus\rep$.
The parameters $k_1-k_8$, $d_0-d_2$ and $c(N,p)$, which are defined in 
(\ref{ks}), (\ref{ds}) and (\ref{Vir}), respectively, are for this
realization
\bes
k_1 &=& k_1(\rep)\dim\,M, \nl
k_2 &=&	k_2(\rep)\dim\,M, \nl
k_3 &=&	k_0(\rep)\dim\,M, \nl
k_4 &=&	\dim\,\rep\dim\,M, \nl
k_5 &=& y_M\dim\,\rep \nl
k_6 &=& z_M\dim\,\rep 
\nlb{kreal}
k_7 &=& z_M k_0(\rep) \nl
k_8 &=& w_M\dim\,\rep, \nl
d_0 &=& \dim\,\rep\dim\,M, \nl
d_1 &=&	k_0(\rep)\dim\,M, \nl
d_2 &=& z_M\dim\,\rep \nl
c(N,p) &=& -\dim\,\rep\dim\,M \Np{}.
\eens
If we substitute these expression into (\ref{cs}), the abelian charges
in Theorems 1 and 3 of \cite{Lar98} are recovered. 

There are some appearent discrepancies. First, we could use a
bosonic Fock representation for (\ref{Freal}), giving all
parameters the opposite sign. Second, the modification (\ref{DF})
leads to the shift (\ref{Dc}) in some of the abelian charges. Third,
in my previous paper I introduced a parameter $w$ which only affects
a trivial cocycle; here I have set $w=\la$.

\section{ Sugawara construction }
\label{sec:Suga}
An immediate corollary of Theorem \ref{T1} is that if we are able 
to construct a represention of
$Vir \ltimes \LUgg$, commuting with
the oscillators $\qmu(t)$ and $\pmu(t)$, we have automatically 
constructed a representation of $DGRO(N)$. As an example we employ the
Sugawara construction to express the reparametrization Virasoro algebra
in terms of the $\LUgg$ generators. The Sugawara construction is of
course well known, and it is described in many places \cite{FMS96,GO86}.
What is somewhat unusual is that $\tr I^M(t)\neq 0$, at least in some
cases, such as $M = {}^{\mm\emp}_\mm$ (no sum on $\mm$). Therefore,
we do not assume that the central charge matrix is proportional to the
unit matrix.
Set
\be
F(t) = \gamma_{MN} \no{ I^M(t) I^N(t) },
\label{Suga}
\ee
where the coefficients $\gamma_{MN} = \gamma_{NM}$ satisfy
\be
\gamma_{MN}(2k^{NR}\dlt^M_T + f^{NR}{}_S f^{SM}{}_T) = \dlt^R_T.
\ee
One finds that the operators (\ref{Suga}) satisfy (\ref{Vir}) --
(\ref{LI}), where
\bes
c(N,p) &=& 2\gamma_{MN}k^{MN}, \nle
d^R&=& i\gamma_{MN} f^{NR}{}_S k^{SM} = 
i\gamma_{MN} f^{MN}{}_S k^{SR} =0,
\eens
and we used (\ref{cond}) in the last step.
In particular, if we assume that
\be
k_{MN} f^{NR}{}_S\ab f^{SM}{}_T = Q\dlt^T_V,
\ee
where $k_{MN}$ is the inverse of $k^{MN}$:
$k^{MN}k_{NR} = k_{RN}k^{NM} = \dlt^M_R$,
then 
\be
\gamma_{MN} = {k_{MN}\/2+Q}, \qquad
c(N,p) = {2\/2+Q} \dlt^M_M. 
\ee
Since $I^M = I^{\mm A}_\nn$,
\be
\dlt^M_M = \summn \dlt^\mm_\nn \dlt^\nn_\mm \dlt^A_A
= \Np{}\dlt^A_A,
\ee
by Lemma \ref{LA} below. 
Clearly, $\dlt^A_A = \dim\,\UU(\gg) = \infty$, so the abelian charge
$c_4$ is infinite, which is unacceptable.
However, the Sugawara construction is well defined if
we replace $\LUgg$ with $\LMgg$, because the underlying associative
algebra is finite-dimensional whenever $M$ is a finite-dimensional
representation.
Hence
\bes
c(N,p) &=& {2\/2+Q} \dim\,M \Np{}, 
\nlb{cSuga}
d_0 &=& d_1 = d_2 = 0.
\eens

We could also consider a hybrid representation,
$F(t) = F_{\oj}(t) + F_{gl(N)}(t)$, where we construct $F_{\oj}(t)$ 
by Sugawara for $\Ug$ and use a Fock representation for
$F_{gl(N)}(t)$.

\section{ A different realization }
\label{sec:long}
It is possible to realize the Virasoro operators $L_f$
in a different fashion. This is based upon the observation that the jet 
$\phim(t)$ in (\ref{jetdef}) can equivalently be defined in terms of 
a Taylor expansion around the observer's trajectory:
\be
\phi(x,t) = \sum_{|\mm|\leq p} {1\/\mm!}\phim(t)(x-q(t))^\mm,
\label{Taylor}
\ee
where $(x-q(t))^\mm = (x^1-q^1(t))^{m_1}\ldots(x^N-q^N(t))^{m_N}$.
The space spanned by $\phi(x,t)$, $x\in\RR^N$, $t\in S^1$, contains
an invariant subspace consisting of $t$-independent fields.
The condition ${\d\/\d t}\phi(x,t)=0$ translates into
\be
\dot\phim(t) = \dot\qmu(t)\phi_{,\mm+\mu}(t),
\label{long}
\ee
for all $\mm$ such that $|\mm|\leq p-1$ (we can not have $|\mm|=p$,
because then $\phi_{,\mm+\mu}(t)$ is undefined). Now substitute this
relation into the expression for $F(t)$ in (\ref{Freal}):
\bes
F(t) &=& \summ \no{\pim(t)\dot\phim(t)}
= \summ \dot\qmu(t)\no{\pim(t)\phi_{,\mm+\mu}(t)} \nle
&=& \dot\qmu(t) \summ E^\mm_{\mm+\mu}(t)
= \dot\qmu(t) P_\mu(t).
\eens

\begin{theorem}
\label{T2}
Embed $E^\mm_\nn(t)$, $J^{\mm a}_\nn(t)$ and $T^{\mm \mu}_{\nn\nu}(t)$
into $\LUgg$ by means of (\ref{EJT}).
Set
\bes
L(t) &=& -\no{\dot\qmu(t)\pmu(t)} + F(t), \nle
F(t) &=& \dot\qmu(t) P_\mu(t),
\eens
where $P_\mu(t) = \sumr E^\rr_{\rr+\mu}(t)$.
Set $\Lxi'=\Lxi+\Delta\Lxi$, $\J_X'=\J_X+\Delta\J_X$, where
\bes
\Delta \Lxi = -{ic_3\/4\pi i}\int dt\ \dmu\xmu(q(t)), 
\nlb{shift}
\Delta \J_X = -{ic_6\/4\pi i}\int dt\ \dlt_a X^a(q(t))
\eens
and $\Lxi$ and $\J_X$ are given in (\ref{DROreal}). 
Then $\Lxi'$, $\J_X'$ and $L_f = \int dt f(t)L(t)$ satisfy
the DGRO algebra $DGRO(N,\oj)$. The abelian charges are the same as in
Theorem \ref{T1}, with the following exceptions:
\bes
c_3 &=&	1 + 2k_3\Np{+1} + 2k_4\Np{+2}, \nl 
c_4 &=& 2N, \\
c_6 &=& 2k_6\Np{+1}.
\eens
\end{theorem}

The proof is deferred to Appendix \ref{appP2}. 

It should be noted that $F(t)$ defined in Theorem \ref{T2} does not
satisfy the relations (\ref{commute}) required by Theorem \ref{T1}, so
this is a genuinely new realization. Instead, (\ref{commute}) is 
replaced by
\be
{[}F(s), \qmu(t)] = 0, \qquad
{[}F(s), \pnu(t)] = P_\nu(s)\dot\dlt(s-t).
\ee

\section{ Finiteness conditions }
\label{sec:finite}
In the previous sections, we constructed realizations of $DGRO(N,\oj)$
in terms of the loop algebra $\LUgg$. If we substitute the realization
(\ref{Freal}) of this loop algebra, we obtain manifestly well defined 
$DGRO(N,\oj)$ Fock modules for each finite $p$.
It is interesting to consider the limit $p\to\infty$. Then the
Taylor expansion (\ref{Taylor}) is essentially equivalent to a
spacetime field $\phi(x,t)$, under some suitable analyticity assumptions.
However, taken at face value, the prospects for taking this limit appear
bleak. When $p$ is large, ${m+p\choose n} \approx p^n/n!$, so 
the abelian charges (\ref{cs}) diverge; the worst case is 
$c_1 \approx c_2 \approx p^{N+2}/(N+2)!$, which diverge in all dimensions
$N > -2$.

There is one way out of this problem. We can consider a more general
realization by taking the direct sum of operators corresponding to 
different values of the jet order $p$. Set therefore
\bes
F(t) &=& F_{(0)}(t) + F_{(1)}(t) + F_{(2)}(t) 
 +\ldots+ F_{(r)}(t), 
\nlb{FIexp}
I^M(t) &=& I^M_{(0)}(t) + I^M_{(1)}(t) + I^M_{(2)}(t) 
 +\ldots+ I^M_{(r)}(t).
\eens
where $F_{(i)}(t)$ and $I^M_{(i)}(t)$ form a basis for 
$Vir\oplus \LUU(\gg,N,p-i)$, and operators corresponding to different
value of the label $(i)$ commute. Thus $(i)$ corresponds to the
jet order $p-i$, and if we only keep one term (i.e. $r=0$), we recover
the situation studied previously.
The equations (\ref{Vir}), (\ref{LI}) and (\ref{KMI}) are replaced by
\bes
[F_{(i)}(s),F_{(j)}(t)] &=& \dlt_{ij} \Big\{
 (F_{(i)}(s)+F_{(i)}(t)) \dot\dlt(s-t) +\nl
&&\qquad+ {c_{(i)}(N,p-i)\/24\pi i}(\dddot\dlt(s-t) 
+ \dot\dlt(s-t))\Big\}, \\
{[}F_{(i)}(s),I^M_{(j)}(t)] &=& \dlt_{ij} \Big\{
 I^M_{(i)}(s)\dot\dlt(s-t)
 + {d^M_{(i)}\/4\pi i}(\ddot\dlt(s-t)+i\dot\dlt(s-t))\Big\}, \nl
{[}I^M_{(i)}(s), I^M_{(j)}(t)] &=& \dlt_{ij} \Big\{
 if^{MN}{}_R I^R_{(i)}(s) \dlt(s-t) 
 + {k^{MN}_{(i)}\/2\pi i}\dot\dlt(s-t)\Big\}.
\eens
It is immediate that (\ref{DROreal}), with these more general expressions
(\ref{FIexp}) for the Virasoro and Kac-Moody currents, still yields a
realization of $DGRO(N,\oj)$.

The abelian charges are given by sum of terms like those in Theorem 
\ref{T1}, e.g.,
\bes
c_5 &=& \sum_{i=0}^r \ki_5 {N+p-i\choose N} 
\nlb{c5f}
&=&\kzero_5{N+p\choose N} + \kone_5{N+p-1\choose N} + \ldots
 + \kr_5{N+p-r\choose N}.
\eens
We now want to choose the parameters $\ki_5$ such that the expression
for $c_5$ has a finite $p\to\infty$ limit when $N\leq r$.
Because of the following lemma, the right choice is
$\ki_5 = \mri k_5$, where $k_5 = \kzero_5$.

\begin{lemma}
\label{P1}
\[
\sum_{i=0}^r \mri {N+p-i\choose N} = \Npr.
\]
\end{lemma}
\bproof
Denote the LHS by $c_{r,p,N}$. Then we use the recurrence formula
in Lemma \ref{L2} to write
\bes
c_{r,p,N} &=& \sum_{i=0}^r (-)^i \Big\{ {r-1\choose i} 
 + {r-1\choose i-1}\Big\} {N+p-i\choose N} \\
&=& \sum_{i=0}^r (-)^i {r-1\choose i} {N+p-i\choose N}
+ \sum_{j=0}^{r-1} (-)^{j+1} {r-1\choose j} {N+p-j-1\choose N}\nl
&=& c_{r-1,p,N} - c_{r-1,p-1,N}.
\eens
But the recurrence formula also implies that
\be
\Npr = {N+p-r+1\choose N-r+1} - {N+p-r\choose N-r+1}.
\qquad\qed
\ee
\eproof

The abelian charges $c_6$ and $c_8$ are given by analogous expressions.
For $c_3$ and $c_7$, the situation is somewhat more complicated, because
there are two terms. To organize calculations, we first introduce the
following function:
\be
G_{r,p,N}(\{\ki\}_{i=0}^r) = \sum_{i=0}^r \ki{N+p-i\choose N}.
\label{G}
\ee

\begin{lemma}
\label{P2}
The function (\ref{G}) has the following properties:
\bes
i.&&G_{r,p,N}(\{\ki\}_{i=0}^r) = \Npr k, 
\hbox{ if $\ki = \mri k$,} \nl
ii.&&G_{r,p,N}(\{\ki\}_{i=0}^r) + G_{r-1,p-1,N}(\{\bar\ki\}_{i=0}^{r-1})
= G_{r,p,N}(\{\ki+\bar k^{(i-1)}\}_{i=0}^r),\nl
&&\hbox{where $\bar k^{(-1)} \equiv 0$.}\nl
iii.&& G_{r,p,N}(\{\ki\}_{i=0}^r) = 
G_{r-1,p,N-1}(\{\sum_{j=0}^i\kj\}_{i=0}^{r-1}),
\hbox{ provided that $\sum_{i=0}^r \ki = 0$.}
\eens
\end{lemma}
\bproof
Property $i$ is equivalent to Lemma \ref{P1}. 
Property $ii$ follows from a
trivial shift in the summation variable in the second term. To prove
property $iii$, 
set $a^{(0)} = \kzero$, $a^{(i)} = a^{(i-1)} + \ki = \sum_{j=0}^i \kj$.
Then
\bes
&&G_{r,p,N}(\{\ki\}_{i=0}^r) =
\kzero{N+p\choose N} + \kone {N+p-1\choose N} + ... + 
\kr {N+p-r\choose N}\nl
&&= a^{(0)}{N+p\choose N} + (a^{(1)}-a^{(0)}) {N+p-1\choose N} + ... + 
(a^{(r)}-a^{(r-1)}) {N+p-r\choose N} \nl
&&= a^{(0)}{N+p-1\choose N-1} + a^{(1)}{N+p-2\choose N-1} + ... +\nl
&&\qquad+ a^{(r-1)}{N+p-r\choose N-1} + a^{(r)}{N+p-r\choose N} \nl
&&= G_{r-1,p,N-1}(\{a^{(i)}\}_{i=0}^{r-1}) + a^{(r)}{N+p-r\choose N},
\eens
by repeated use of the recurrence formula.
\qed
\eproof

Using the function $G_{r,p,N}$ defined in (\ref{G}), we rewrite the
expression for $c_7$ as
\bes
c_7 &=& \sum_{i=0}^r \ki_7 {N+p-i\choose N} +
\sum_{i=0}^r \ki_6 {N+p-i\choose N+1} \nl
&=& G_{r,p,N}(\{\ki_7\}_{i=0}^r) +
G_{r,p-1,N+1}(\{\ki_6\}_{i=0}^r) \nl
&=& G_{r,p,N}(\{\ki_7\}_{i=0}^r) +
 G_{r-1,p-1,N}(\{\sum_{j=0}^i\kj_6\}_{i=0}^{r-1}) \\
&=& G_{r,p,N}(\{\ki_7+\sum_{j=0}^i\kj_6\}_{i=0}^r) \nl
&=& k_7\Npr,
\eens
which holds provided that $\sum_{i=0}^r \ki_6 = 0$ and
$\ki_7+\sum_{j=0}^{i-1}\kj_6 = \mri k_7$. The calculation
made use of all three properties in Lemma \ref{P2}.
The abelian charge $c_3$ is treated analogously. 

For $c_1$, we must apply Lemma \ref{P2} twice to the last term:
\bes
1-c_1 &=& \sum_{i=0}^r \ki_1 {N+p-i\choose N}
+ \sum_{i=0}^r \ki_4 {N+p-i+1\choose N+2} \nl
&=& G_{r,p,N}(\{\ki_1\}_{i=0}^r) +
G_{r,p-1,N+2}(\{\ki_4\}_{i=0}^r) \nl
&=& G_{r,p,N}(\{\ki_1\}_{i=0}^r) +
G_{r-1,p-1,N+1}(\{\sum_{j=0}^i\kj_4\}_{i=0}^{r-1}) \\
&=& G_{r,p,N}(\{\ki_1\}_{i=0}^r) +
G_{r-2,p-1,N}(\{\sum_{j=0}^i\sum_{\ell=0}^j \kl_4\}_{i=0}^{r-2}) \nl
&=& G_{r,p,N}(\{\ki_1 + \sum_{j=0}^{i-1}\sum_{\ell=0}^j \kl_4\}_{i=0}^r) \nl
&=& k_1 \Npr,
\eens
where the conditions are
\bes
\sum_{i=0}^r \ki_4 &=& 0, \nl
\sum_{i=0}^{r-1}\sum_{j=0}^i \kj_4 &=& 0, \\
\ki_1 + \sum_{j=0}^{i-1}\sum_{\ell=0}^j \kl_4 &=& \mri k_1.
\eens
$c_2$ is computed by repeatedly using the properties in
Lemma \ref{P2}:
\bes
-c_2 &=& \sum_{i=0}^r \ki_2 {N+p-i\choose N}
+ \sum_{i=0}^r 2\ki_3 {N+p-i\choose N+1} 
+ \sum_{i=0}^r \ki_4 {N+p-i\choose N+2} \nl
&=& G_{r,p,N}(\{\ki_2\}_{i=0}^r) +
G_{r,p-1,N+1}(\{2\ki_3\}_{i=0}^r) +
G_{r,p-2,N+2}(\{\ki_4\}_{i=0}^r) \nl
&=& G_{r,p,N}(\{\ki_2\}_{i=0}^r) +
G_{r,p-1,N+1}(\{2\ki_3\}_{i=0}^r) +
G_{r-1,p-2,N+1}(\{\bar\ki_4\}_{i=0}^{r-1}) \nl
&=& G_{r,p,N}(\{\ki_2\}_{i=0}^r) +
G_{r,p-1,N+1}(\{2\ki_3+\bar k^{(i-1)}_4\}_{i=0}^r) \nl
&=& G_{r,p,N}(\{\ki_1 + \sum_{j=0}^{i-1}(2\ki_3+\bar k^{(i-1)}_4)
 \}_{i=0}^r) \\
&=& k_1 \Npr,
\eens
where $\bar\ki_4 = \sum_{j=0}^i\kj_4$,
and the conditions are
\bes
\bar\kr_4 &=& 0, \nl
\sum_{i=0}^r (2\ki_3 + \bar k^{(i-1)}_4) &=& 0, \nle
\ki_2 + \sum_{j=0}^{i-1}(2\kj_3 + \bar k^{(j-1)}_4) &=& \mri k_2.
\eens

Finally, the Virasoro central charge $c(N,p)$, when given by either 
(\ref{kreal}) or (\ref{cSuga}), has the form $c(N,p) = c\Np{}$ for some
parameter $c$ independent of both $N$ and $p$. 
The abelian charge $c_4$ then has the form $c_4 = 2N+c\Np{}$. When we
add several contributions $c^{(i)}$, it becomes
\be
c_4 = 2N + \Npr,
\label{finc4} 
\ee
provided that $c^{(i)} = \mri c$. 

The results in this section can be summarized in the following theorem:

\begin{theorem}
\label{T3}
Let the operators $F(t)$ and $I^M(t)$ be given as in (\ref{FIexp}), and
let $E^\mm_\nn(t)$, $J^{\mm a}_\nn(t)$, $T^{\mm \mu}_{\nn\nu}(t)$ be as
in (\ref{EJT}).
Then the generators (\ref{DROreal}) satisfy the DGRO algebra
$DGRO(N,\oj)$ (\ref{DGRO}). The abelian charges are given by
\bes
c_1 &=& 1 - k_1\Npr, \nl
c_2 &=& -k_2\Npr, \nl
c_3 &=& 1 + d_1\Npr, \nl
c_4 &=& 2N + c\Npr,
\nlb{finc} 
c_5 &=& k_5\Npr, \nl
c_6 &=& d_2\Npr, \nl
c_7 &=& k_7\Npr, \nl
c_8 &=& k_8\Npr,
\eens
where $\kzero_n = k_n$, $\dzero_n = d_n$, provided that the following
conditions hold:
\bes
&&\ki_1 + \sum_{j=0}^{i+1}\sum_{\ell=0}^j \kl_4 = \mri k_1, \nl
&&\ki_2 + \sum_{j=0}^{i-1}(2\kj_3 + \sum_{\ell=0}^{j-1} \kl_4) 
 = \mri k_2, \nl
&&\sum_{i=0}^r (2\ki_3 + \sum_{\ell=0}^{i-1} \kl_4) = 0, \nl
&&\sum_{i=0}^r \ki_4 = 0, \nl
&&\sum_{i=0}^{r-1}\sum_{j=0}^i \kj_4 = 0, \nl
&&\ki_5 = \mri k_5, \\
&&\sum_{i=0}^r \ki_6 = 0, \nl
&&\ki_7+\sum_{j=0}^{i-1}\kj_6 = \mri k_7, \nl
&&\ki_8 = \mri k_8, \nl
&&\sum_{i=0}^r \di_0 = 0, \nl
&&\di_1+\sum_{j=0}^{i-1}d^{(j)}_0 = \mri d_1, \nl
&&\di_2 = \mri d_2, \nl
&&c^{(i)} = \mri c.
\eens
All abelian charges vanish if $N<r$ and diverge when $p\to\infty$ if 
$N>r$. When $N=r$, the abelian charges are independent of $p$ and in 
general non-zero.
\end{theorem}

\section{Discussion}
In this paper, I reformulated the $DGRO(N,\oj)$ Fock modules 
from \cite{Lar98} as realizations in terms of Virasoro and affine
Kac-Moody
currents. This gave rise two new types of realizations: the Sugawara
construction in Section \ref{sec:Suga} and the modification in 
Section \ref{sec:long}, and it became possible to formulate conditions
for the existence of the $p\to\infty$ limit.

The modules obtained in the last section are quite unnatural. True, they
admit that we let $p\to\infty$, but in the same time they are highly 
reducible, being direct sums for finite $p$. So we are in the strange 
situation that
there exist several well-defined modules for finite $p$, but only their
sum and not the individual summands survive in the limit. Another problem
is that the central charges of the underlying affine algebras have 
alternating signs, by Theorem \ref{T3}, so they can not all be represented
unitarily.

There is an attractive resolution to these problems. If we can find an
invariant nilpotent fermionic operator intertwining between the
individual Fock modules in the direct sum,
the associated cohomology groups will also be modules. The modules at
jet order $p$ can be thought of as quantum fields, whereas lower-order
modules describe antifields of various degree. In the unpublished paper
\cite{Lar99}, I began to investigate a cohomology theory closely related
to the Koszul-Tate cohomology arising in the Batalin-Vilkovisky approach
to gauge theory \cite{HT92}. Note that the better known BRST cohomology
does not work, for two reasons. First, BRST both imposes
constraints and identifies points on gauge orbits, which means that the
gauge algebra acts trivially on the cohomology. Second, BRST is ill 
defined in the presence of non-trivial cocycles, which is precisely 
the case of interest here.

The conditions in Theorem \ref{T3} impose severe restrictions on
the possible field content and on the form of the Euler-Lagrange equations.
Preliminary calculations indicate that if we assume that the 
Euler-Lagrange
equations are of first order for fermions and second order for bosons,
and we only have irreducible gauge symmetries of one order higher, then
the dimension of space-time must be four. Moreover, it is also necessary
to have gauge symmetries for fermions, pointing toward some kind of
supersymmetry. These issues will be addressed in a forthcoming publication.

%\section{Appendices}

\appendix

\section{ Lemmas }
\label{sec:A}

\begin{lemma} \label{L0}
\[
{\mm\choose\nn} {\nn\choose\rr} = 
{\mm\choose\rr} {\mm-\rr\choose\mm-\nn}.
\]
\end{lemma}

\bproof
Both sides can be written as
\[
{\mm!\/\nn!(\mm-\nn)!} {\nn!\/\rr!(\nn-\rr)!} {(\mm-\rr)!\/(\mm-\rr)!}.
\qquad\qed
\]
\eproof

\begin{lemma} \label{L1}
\[
\summnr {\mm\choose\nn} {\nn\choose\rr}
\d_{\mm-\nn} f \d_{\nn-\rr} g
= \summr {\mm\choose\rr} \d_{\mm-\rr} (fg).
\]
\end{lemma}

\bproof 
\bes
LHS &=& \summnr {\mm\choose\rr} {\mm-\rr\choose\mm-\nn}
\d_{\mm-\nn} f \d_{\nn-\rr} g \nl
&=& \summr {\mm\choose\rr}\sums {\mm-\rr\choose\ss}
\d_\ss f \d_{\mm-\rr-\ss}g
=RHS,
\eens
by Lemma \ref{L0} and Leibniz' rule. 
\qed
\eproof 

\begin{lemma} \label{L2}
\[
{\mm\choose\nn} + {\mm\choose\nn-\mu} - {\mm+\mu\choose\nn} = 0.
\]
In one dimension, this is the {\em recurrence formula}:
\[
{m+1\choose n} = {m\choose n} + {m\choose n-1}.
\]
\end{lemma}

\bproof
The three terms only differ in the $\mu\th$ components, so the other
components contribute a constant factor. It thus suffices to prove the
one-dimensional recurrence formula:
\[
{m\choose n} + {m\choose n-1} - {m+1\choose n} 
= {m\choose n} (1 + {n\/m-n+1} - {m+1\/m+1-n}) = 0.
\qquad\qed
\]
\eproof 

\begin{lemma} \label{L3}
\[
{\mm\choose\nn}{\nn+\mu\choose\ss} =
{\mm\choose\ss}{\mm-\ss\choose\mm-\nn}
+{\mm\choose\nn}{\nn\choose\ss-\mu}.
\]
\end{lemma}

\bproof 
First use Lemma \ref{L2} on ${\nn+\mu\choose\ss}$, then apply 
Lemma \ref{L0}.
\qed
\eproof

\begin{lemma}
\label{LA}
\[
\summp{} 1 = \Np{}.
\]
\end{lemma}

\bproof 
The statement is proved by induction. Define
\bes
A(N,p) &\equiv& \summp{} 1 
= \sum_{m_1=0}^p\dsum{m_2+...+m_N}{\leq p-m_1} 1 
\nlb{Adef}
&=& \sum_{m_1=0}^p A(N-1, p-m_1).
\eens
Thus,
\bes
A(N,p+1) &=& \sum_{m=0}^{p+1} A(N-1, p+1-m) \nle
&=& \sum_{n=0}^p A(N-1,p-n) + A(N-1,p+1),
\eens
where we substituted $n=m-1$. $A(N,p)$ must therefore obey the recursion 
relation
\be
A(N,p+1) = A(N,p) + A(N-1,p+1).
\label{Arec}
\ee
The equation has the solution 
\be
A(N,p)=\Np{}
\label{A}
\ee
due to the recurrence relation in Lemma \ref{L2}.
Moreover, the boundary case $p=0$ is clear, since 
\[
A(N,0) = \sum_{|\mm|=0} 1 = 1 = {N\choose N}.
\qquad\qed
\]
\eproof 

\begin{lemma}
\label{LB}
\[
\sum_{\mu}\summp{-1}{\mm+\mu\choose\mm}\phi^\mu_\mu 
= \Np{+1}\phi^\mu_\mu.
\]
\end{lemma}

\bproof 
By symmetry, we only need to evaluate the expression for $\mu=1$.
Set
\be
B(N,p)\equiv \summp{}{\mm+\one\choose\mm} = \summp{} (m_1+1),
\ee
where $\one$ denotes a unit vector in the first direction.
Clearly,
\be
B(N,p) = \sum_{m_1=0}^p\dsum{m_2+...+m_N}{\leq p-m_1} (m_1+1) 
= \sum_{m=0}^p (m+1) A(N-1, p-m).
\label{Bdef}
\ee
This function satisfies
\bes
B(N,p+1) &=& \sum_{m=0}^{p+1} (m+1) A(N-1, p+1-m) \nl
&=& \sum_{n=0}^p (n+2) A(N-1, p-n) + A(N-1, p+1) \nl
&=& B(N,p) + A(N,p) + A(N-1, p+1)
\label{Brec} \\
&=& B(N,p) + A(N,p+1) \nl
&=& B(N,p) + {N+p+1\choose N}.
\eens
By Lemma \ref{L2}, this recursion relation has the solution
\be
B(N,p) = {N+p+1\choose N+1}.
\label{B}
\ee
The boundary case is clear, because
\[
B(N,0) = \sum_{|\mm|=0} (m_1+1) = 1 = {N+1\choose N+1}.
\]
Thus,
\[
\summp{-1}{\mm+\one\choose\mm}\phi^1_1
= B(N,p-1)\phi^1_1 = \Np{+1}\phi^1_1,
\]
and the lemma is proven. \qed
\eproof

\begin{lemma}
\label{LC}
\bes
&&\sum_{\mu,\nu}
\summp{-1}{\mm+\mu\choose\mm}{\mm+\nu\choose\mm} \phi^{\mu\nu}_{\nu\mu}
+ \sum_{\mu\neq\nu}
\sumrp{-2}{\rr+\mu+\nu\choose\rr+\nu}{\rr+\mu+\nu\choose\rr+\mu}
\phi^{\mu\nu}_{\mu\nu} \nl
&&= {N+p+1\choose N+2} \phi^{\mu\nu}_{\nu\mu}
+ {N+p\choose N+2} \phi^{\mu\nu}_{\mu\nu}
\label{Clemma}
\ees
\end{lemma}

\bproof 
We have to distinguish two cases: $\mu\neq\nu$ and $\mu=\nu$.
Consider the first case, say $\mu=1$, $\nu=2$, and define
\bes
C(N,p) &=& \summp{} {\mm+\one\choose\mm}{\mm+\two\choose\mm}
= \summp{} (m_1+1)(m_2+1) \nl
&=& \sum_{m_1=0}^p \sum_{m_2=0}^{p-m_1} 
\dsum{m_3+...+m_N}{\leq p-m_1-m_2}(m_1+1)(m_2+1) 
\label{Cdef}\\
&=&\sum_{m_1=0}^p \sum_{m_2=0}^{p-m_1}(m_1+1)(m_2+1) A(N-2,p-m_1-m_2).
\eens
The recursion relation becomes
\bes
C(N,p+1) &=& \sum_{m_1=0}^{p+1} \sum_{m_2=0}^{p+1-m_1}
 (m_1+1)(m_2+1) A(N-2,p+1-m_1-m_2) \nl
&=& \sum_{n_1=0}^p \sum_{m_2=0}^{p-n_1}(n_1+2)(m_2+1) A(N-2,p-n_1-m_2) \nl
&&+ \sum_{m_2=0}^{p+1}(m_2+1) A(N-2,p+1-m_2) 
\label{Crec1} \\
&=& C(N,p) + \sum_{n_1=0}^p B(N-1, p-n_1) + B(N-1,p+1),
\eens
where we used (\ref{Bdef}) twice. We now need the sum
\be
\Delta(N,p) = \sum_{m=0}^p B(N-1,p-m) = \sum_{m=0}^p {N+p-m\choose N}.
\ee
Recursion relation:
\bes
\Delta(N,p+1) &=& \sum_{m=0}^{p+1} {N+p+1-m\choose N} \nl
&=& \sum_{n=0}^p {N+p-n\choose N} + {N+p+1\choose N} \\
&=& \Delta(N,p) + {N+p+1\choose N},
\eens
with the solution 
\be
\Delta(N,p) = {N+p+1\choose N+1}.
\label{Delta}
\ee
We now substitute (\ref{Delta}) into (\ref{Crec1}):
\bes
C(N,p+1) &=& C(N,p) + {N+p+1\choose N+1} + {N+p+1\choose N} \nl
&=& C(N,p) + {N+p+2\choose N+1} 
\label{Crec}\\
&=& C(N,p) + B(N,p+1),
\eens
and thus
\be
C(N,p) = {N+p+2\choose N+2}.
\label{C}
\ee
We check that the boundary value $C(N,0)=1$ is correct.
The $\mu=1$, $\nu=2$, contribution to the first term in (\ref{Clemma})
is thus $C(N,p-1) \phi^{12}_{21}$.
To obtain the contribution to the second term, we note that
\bes
&&\sumrp{-2}{\rr+\one+\two\choose\rr+\two}{\rr+\one+\two\choose\rr+\one} \nl
&&= \sumrp{-2} (r_1+1)(r_2+1) = C(N,p-2).
\eens
Hence the total contribution to (\ref{Clemma}) is
\be
C(N,p-1) \phi^{12}_{21} + C(N,p-2)\phi^{12}_{12}.
\label{C12}
\ee

Now we consider the case $\mu=\nu=1$, say. The relevant sum is
\be
D(N,p) = \summp{} (m_1+1)^2 = \sum_{m=0}^p (m+1)^2 A(N-1,p-m).
\label{Ddef}
\ee
Recursion relation:
\bes
D(N,p+1) &=& \sum_{m=0}^{p+1} (m+1)^2 A(N-1,p+1-m) \nl
&=& \sum_{n=0}^p ((n+1)^2 + 2(n+1) + 1) A(N-1,p-n) + A(N-1,p+1) \nl
&=& D(N,p) + 2B(N,p) + A(N,p) + A(N-1,p+1) 
\label{Drec} \\
&=& D(N,p) + B(N,p) + B(N,p+1),
\eens
where (\ref{Brec}) was used in the last step.
Equation (\ref{Drec}) has the solution
\[
D(N,p) = C(N,p) + C(N,p-1) = {N+p+2\choose N+2} + {N+p+1\choose N+2},
\]
because substitution of this expression into (\ref{Drec}) gives rise to 
two copies of the identity (\ref{Crec}) (for $p$ and $p-1$, respectively).
The total contribution to the LHS is thus
\[
(C(N,p-1) + C(N,p-2))\phi^{11}_{11}
= C(N,p-1)\phi^{11}_{11} + C(N,p-2)\phi^{11}_{11},
\]
which is of the same form as (\ref{C12}).
\qed
\eproof 

\begin{lemma}
\label{LE}
\bes
\sum_{\mu,\nu} \sumnp{-2} {\mm+\mu+\nu\choose\mm} \phi^{\mu\nu}_{\mu\nu} =
\Np{+2} \phi^{\mu\nu}_{\mu\nu}.
\eens
\end{lemma}

\bproof
First consider the case $\mu=\nu=1$, say. The relevant sum is
\bes
E(N,p) &=& 2\summp{} {\mm+2\one\choose\mm} \nle
&=& \sum_{m_1=0}^p (m_1+2)(m_1+1) A(N-1,p-m_1).
\eens
Recursion relation:
\bes
E(N,p+1) &=& \sum_{m=0}^{p+1} (m+2)(m+1) A(N-1,p+1-m) \nl
&=& \sum_{n=0}^p (n+3)(n+2) A(N-1,p-n) + 2A(N-1,p+1) \nl
&=& E(N,p) + 2 \sum_{n=0}^p (n+2)A(N-1,p-n)+ 2A(N-1,p+1) \nl
&=& E(N,p) + 2 B(N,p+1) = E(N,p) + 2{N+p+2\choose N+1}.
\eens
Hence
\be
E(N,p) = 2{N+p+2\choose N+2}
\ee
and
\be
\sumnp{-2} {\mm+\one+\one\choose\mm} \phi^{11}_{11}
= \half E(N,p-2) \phi^{11}_{11}
= \Np{+2}\phi^{11}_{11}.
\label{E1}
\ee
Now let $\mu\neq\nu$, say $\mu=1$, $\nu=2$.
\be
\sumnp{-2} {\mm+\one+\two\choose\mm} \phi^{12}_{12}
= C(N,p-2)\phi^{12}_{12}
= \Np{+2}\phi^{12}_{12}.
\label{E2}
\ee
The lemma now follows by summing over contributions of the form 
(\ref{E1}) and (\ref{E2}).
\qed
\eproof

\begin{lemma}
\label{Lemb}
There is an embedding
$diff(N) \into diff(N)\ltimes map(N,gl(N))$, given by
$\Lxi' = \Lxi+T_{d\xi}$.
\end{lemma}

\bproof
It suffices to prove the statement for the realization
$\Lxi = \xmu\dmu$, $T_{d\xi} = \dnu\xmu T^\nu_\mu$, where
$T^\mu_\nu$ provide a realization of $gl(N)$ (\ref{glN}):
\bes
[\Lxi',\Leta'] &=&
[\xmu\dmu + \drho\xmu T^\rho_\mu, \ynu\dnu + \dsi\ynu T^\si_\nu] \nl
&=& \xmu\dmu\ynu\dnu + \xmu\dmu\dsi\ynu T^\si_\nu
+ \drho\xmu \dsi\ynu \dlt_\mu^\si T^\rho_\nu - \xxy \nl
&=& \xmu\dmu\ynu\dnu + \dsi(\xmu\dmu\ynu) T^\si_\nu - \xxy \nl
&=& [\xi,\eta]^\nu\dnu + \dsi[\xi,\eta]^\nu T^\si_\nu
= \L'_{[\xi,\eta]}.
\qquad\qed
\eens
\eproof

\section{ Proof of Theorem \ref{T1} }
\label{sec:Pf1}

\subsection{ $\J\J$ bracket }
\label{ssec:JJ}
To reduce writing, we suppress all arguments in single and double
integrals throughout all proofs. Thus we write
$X_a = X_a(q(s))$, $Y_b = Y_b(q(t))$, $J^{\mm a}_\nn = J^{\mm a}_\nn(s)$,
$J^{\rr b}_\ss = J^{\rr b}_\ss(t)$, $\dlt = \dlt(s-t)$ and 
$\dot\dlt = \dot\dlt(s-t) = -\dot\dlt(t-s)$.
Further, $\iint F = \iint dsdt\ F(s,t)$ for any functional $F$, etc.
The representation condition for the $\J\J$ bracket reads
\bes
[\J_X, \J_Y] &=&
\summn \sumrs {\mm\choose\nn} {\rr\choose\ss} \iint
[\d_{\mm-\nn} X_aJ^{\mm a}_\nn, \d_{\rr-\ss} Y_bJ^{\rr b}_\ss] \nl
&=& \summnrs {\mm\choose\nn} {\rr\choose\ss} \iint
\d_{\mm-\nn} X_a \d_{\rr-\ss} Y_b \times\nl
&&\times (\dlt^\rr_\nn g^{ab}{}_C J^{\mm C}_\ss \dlt
- \dlt^\mm_\ss g^{ba}{}_C J^{\rr C}_\nn \dlt
+ {k^{ab}\/2\pi i} \dlt^\mm_\ss \dlt^\rr_\nn \dot\dlt) \\
&=& \summns g^{ab}{}_C \int {\mm\choose\nn} {\nn\choose\ss}
 \d_{\mm-\nn} X_a \d_{\nn-\ss} Y_b J^{\mm C}_\ss \nl
&&- \summnr g^{ba}{}_C \int {\mm\choose\nn} {\rr\choose\mm}
 \d_{\mm-\nn} X_a \d_{\rr-\mm} Y_b J^{\rr C}_\nn \nl
&&-{k^{ab}\/2\pi i}\summnp{} {\mm\choose\nn} {\nn\choose\mm} 
\int \d_{\mm-\nn} \dot X_a \d_{\nn-\mm} Y_b.
\eens
We now use Lemma \ref{L1} to rewrite the first two terms as
\be
\summn {\mm\choose\nn}(g^{ab}{}_C - g^{ba}{}_C) \int
\d_{\mm-\nn}(X_a Y_b)J^{\mm C}_\nn = \J_{[X,Y]}.
\ee
The third term contains the factor ${\mm\choose\nn} {\nn\choose\mm}$,
which is zero unless $\mm = \nn$ (if $m_\rho > n_\rho$ for some
$\rho$, ${\nn\choose\mm}=0$, and if $n_\rho > m_\rho$, 
${\mm\choose\nn}=0$). Therefore, this term becomes
\be
-{k^{ab}\/2\pi i} \summp{} 1  \int \dot X_a Y_b 
= -{k_5\dlt^{ab}+k_8\dlt^a\dlt^b\/2\pi i} \Np{} \int \dot X_a Y_b,
\label{dXY}
\ee
by Lemma \ref{LA}. This fixes $c_5 = k_5\Np{}$ and $c_8 = k_8\Np{}$.

\subsection{ $\L\J$ bracket }
Consider first the case $T_{d\xi} = 0$.
Moreover, we initially ignore normal ordering,
to get the regular terms:
\bes
[\Lxi, \J_X] &=& \iint [\xmu(\pmu-P_\mu)
 + \summn {\mm\choose\nn}\d_{\mm-\nn}\xmu E^\mm_{\nn+\mu}, 
 \sumrs {\rr\choose\ss}\d_{\rr-\ss}X_a J^{\rr a}_\ss] \nl
&=& \int \sumrs {\rr\choose\ss} \Big\{
\xmu \d_{\rr-\ss+\mu}X_a J^{\rr a}_\ss
+ \xmu \d_{\rr-\ss}X_a (J^{\rr a}_{\ss+\mu} - J^{\rr-\mu a}_\ss) \nl
&&+ \summn {\mm\choose\nn}\d_{\mm-\nn}\xmu\d_{\rr-\ss}X_a
(\dlt^\rr_{\nn+\mu} J^{\mm a}_\ss - \dlt^\mm_\ss J^{\rr a}_{\nn+\mu} )
\Big\}\\
&=& \int  \sumrs ({\rr\choose\ss}+{\rr\choose\ss-\mu}-{\rr+\mu\choose\ss})
 \xmu\d_{\rr-\ss+\mu}X_a J^{\rr a}_\ss \nl
&&+\summns {\mm\choose\nn}{\nn+\mu\choose\ss}
 \d_{\mm-\nn}\xmu \d_{\nn+\mu-\ss} X_a J^{\mm a}_\ss \nl
&&-\summnr {\mm\choose\nn}{\rr\choose\mm}
 \d_{\mm-\nn}\xmu \d_{\rr-\mm} X_a J^{\rr a}_{\nn+\mu}.
\eens
The first term vanishes due to Lemma \ref{L1}. In the second, we
use Lemma \ref{L2} and obtain
\bes
&&\summns {\mm\choose\nn}({\nn\choose\ss-\mu}+{\nn\choose\ss})
 \d_{\mm-\nn}\xmu \d_{\nn+\mu-\ss} X_a J^{\mm a}_\ss -\nl
&&-\summnr {\mm\choose\nn}{\rr\choose\mm} 
 \d_{\mm-\nn}\xmu \d_{\rr-\mm} X_a J^{\rr a}_{\nn+\mu} \\
&=& \sums {\mm\choose\ss-\mu}\d_{\mm-\ss+\mu} (\xmu X_a) J^{\mm a}_\ss +\nl
&&+ \sums {\mm\choose\ss}\d_{\mm-\ss} (\xmu \dmu X_a) J^{\mm a}_\ss
- \sumnr {\rr\choose\nn}\d_{\rr-\nn} (\xmu X_a) J^{\rr a}_{\nn+\mu}.
\eens
After we set $\ss=\nn+\mu$ in the first term, it cancels the last term
and we are left with the middle term, which equals $\J_{\xi X}$.

The extension becomes
\bes
&&\ext([\Lxi, \J_X]) = \iint {k^{a\emp}\/2\pi i}\Big\{
- \sumrs {\rr\choose\ss}\xmu\d_{\rr-\ss}X_a \dlt^\rr_{\ss+\mu} +\nl
&&\qquad+ \summnrs {\mm\choose\nn}{\rr\choose\ss}
\d_{\mm-\nn}\xmu\d_{\rr-\ss}X_a \dlt^\mm_\ss\dlt^\rr_{\nn+\mu}
\Big\} \dot\dlt 
\label{appLJ}\\
&&= {k^{a\emp}\/2\pi i}\int \Big\{
\sumsp{-1} {\ss+\mu\choose\ss}\dot\xmu\dmu X_a
- \sums\sumnp{-1} {\ss\choose\nn}{\nn+\mu\choose\ss} \d_{\ss-\nn}\dot\xmu
\d_{\nn-\ss+\mu}X_a \Big\}.
\eens
The last term is zero unless $n_\mu \leq s_\mu$, $s_\mu \leq n_\mu+1$,
and $n_\nu = s_\nu$ for all $\nu\neq\mu$. This leaves two cases:
\begin{alignat}{2}
1.\ &\ss=\nn:\ & -\sumnp{-1} {\nn+\mu\choose\nn} \dot\xmu \dmu X_a, \nle
2.\ &\ss=\nn+\mu:\ & -\sumnp{-1} {\nn+\mu\choose\nn} \dmu\dot\xmu X_a.
\nonumber
\end{alignat}
The first term cancels the first term in (\ref{appLJ}), so the total
extension reads
\be
-{k^{a\emp}\/2\pi i}\int \sum_\mu\sumnp{-1} 
{\nn+\mu\choose\nn} \dmu\dot\xmu X_a
= -{k_6\dlt^a\/2\pi i} \Np{+1} \int \dmu\dot\xmu X_a,
\label{xx1}
\ee
by Lemma \ref{LB}. 
Now replace $\Lxi \mapsto \Lxi + T_{d\xi}$ and use the result in
subsection \ref{ssec:JJ}.
The regular terms vanish because $[d\xi,X] = 0$, and the extension is
given by (\ref{dXY}):
\be
[T_{d\xi}, J_X] = -{k^{\nu a}_\mu \/2\pi i}\Np{} \int \dnu\dot\xmu X_a.
\label{xx2}
\ee
Summing (\ref{xx1}) and (\ref{xx2}), we see that
$c_7 = k_6\Np{+1} + k_7\Np{}$.

\subsection{$\L\L$ bracket}
Again we initially set $T_{d\xi}=0$:
\bes
[\Lxi,\Leta] &=& \iint
[\no{\xmu\pmu} - \xmu P_\mu 
 + \summn {\mm\choose\nn}\d_{\mm-\nn}\xmu E^\mm_{\nn+\mu},
\\
&&\qquad \no{\ynu\pnu} - \ynu P_\nu
 + \sumrs {\rr\choose\ss}\d_{\rr-\ss}\ynu E^\rr_{\ss+\nu}].
\eens
We first ignore normal ordering and extensions, and verify that
the regular terms come out right:
\bes
&& \int \xmu\dmu\ynu\pnu - \xmu\dmu\ynu P_\nu 
+ \sumrs {\rr\choose\ss}\xmu\d_{\rr-\ss+\mu}\ynu E^\rr_{\ss+\nu} \nl
&&- \sumrs {\rr\choose\ss}\xmu\d_{\rr-\ss}\ynu 
 (E^{\rr-\mu}_{\ss+\nu} - E^\rr_{\ss+\nu+\mu}) \nl
&&+\summnrs {\mm\choose\nn}{\rr\choose\ss} 
 \d_{\mm-\nn}\xmu\d_{\rr-\ss}\ynu \dlt^\rr_{\nn+\mu} E^\mm_{\ss+\nu}
 - \xxy \\
&=& \int \xmu\dmu\ynu(\pnu - P_\nu)
 + \sumrs ({\rr\choose\ss} - {\rr+\mu\choose\ss} + {\rr\choose\ss-\mu} )
  \xmu\d_{\rr-\ss+\mu}\ynu E^\rr_{\ss+\nu} \nl
&&+\summns {\mm\choose\nn}{\nn+\nu\choose\ss}
 \d_{\mm-\nn}\xmu\d_{\nn+\mu-\ss}\ynu E^\mm_{\ss+\nu} - \xxy,
\eens
where $\xxy$ stands for terms obtained by interchanging $\xi$ and
$\eta$ everywhere.
The first sum vanishes due to Lemma \ref{L2}, whereas we use 
Lemma \ref{L3} to rewrite the last sum as
\be
\summns ( {\mm\choose\ss}{\mm-\ss\choose\mm-\nn}
+ {\mm\choose\nn}{\nn\choose\ss-\mu})
 \d_{\mm-\nn}\xmu\d_{\nn+\mu-\ss}\ynu E^\mm_{\ss+\nu}.
\ee
In the first term, we set $\rr=\mm-\nn$ and use Leibniz' rule:
\bes
\summrs {\mm\choose\ss}{\mm-\ss\choose\rr}
\d_\rr\xmu \d_{\mm+\mu-\ss-\rr}\ynu E^\mm_{\ss+\nu}
=\summs	{\mm\choose\ss}\d_{\mm-\ss}(\xmu\dmu\ynu) E^\mm_{\ss+\nu}.
\eens
In the second term, we set $\rr=\ss-\mu$, and get
\[
\summnr {\mm\choose\nn}{\nn\choose\rr}
 \d_{\mm-\nn}\xmu\d_{\nn-\rr}\ynu E^\mm_{\rr+\mu+\nu}
= \summr {\mm\choose\rr} \d_{\mm-\rr}(\xmu\ynu)E^\mm_{\rr+\mu+\nu}.
\]
This term is symmetric under the interchange $\xxy$, so
it cancels one of the unwritten terms.
Summing up, we find that the regular terms combine to
\[
\int \xmu\dmu\ynu(\pnu - P_\nu)
 + \summs {\mm\choose\ss}\d_{\mm-\ss}(\xmu\dmu\ynu) E^\mm_{\ss+\nu} - \xxy
= \L_{[\xi,\eta]}.
\]

We now turn to the extensions. The current-independent term was 
calculated in \cite{Lar98} (and, using a different formalism, by
Rao and Moody \cite{RM94}). The result is
\be
\ext([\int\! \no{\xmu\pmu}, \int \! \no{\ynu\pnu}])
=\tpi\int \dnu\dot\xmu\dmu\ynu.
\label{ext0}
\ee
The rest of the extension becomes
\bes
&&\ktpi \iint \Big\{
-\sumrs {\rr\choose\ss}\xmu\d_{\rr-\ss}\ynu \dlt^\rr_{\ss+\nu+\mu} 
-\summn {\mm\choose\nn}\d_{\mm-\nn}\xmu\ynu\dlt^\mm_{\nn+\mu+\nu} \nl
&&+\summnrs {\mm\choose\nn} {\rr\choose\ss}
\d_{\mm-\nn}\xmu \d_{\rr-\ss}\ynu \dlt^\mm_{\ss+\nu} \dlt^\rr_{\nn+\mu}
\Big\} \dot\dlt 
\label{appLL}\\
&=&\ktpi \int \Big\{
\sumsp{-2} {\ss+\mu+\nu\choose\ss}\dot\xmu\d_{\mu+\nu}\ynu 
+\sumnp{-2} {\nn+\mu+\nu\choose\nn}\d_{\mu+\nu}\dot\xmu\ynu -\nl
&&-\sumnsp{-1} {\ss+\nu\choose\nn} {\nn+\mu\choose\ss}
\d_{\ss-\nn+\nu}\dot\xmu \d_{\nn-\ss+\mu}\ynu
\Big\}.
\eens
In the last term, there are two possibilities. First assume that 
$\mu\neq\nu$. Then ${\ss+\nu\choose\nn}$ vanishes unless
$n_\nu\leq s_\nu+1$, $n_\mu\leq s_\mu$,	and ${\nn+\mu\choose\ss}$
vanishes unless $s_\mu\leq n_\mu+1$, $s_\nu\leq s_\nu$. In addition,
both binomial coefficients vanish unless
$n_\rho=s_\rho$ for all $\rho\neq\mu,\nu$. This leaves four
non-zero cases:
\begin{alignat}{2}
1.\ &\nn=\ss:\ & -\sumnp{-1} {\nn+\nu\choose\nn} {\nn+\mu\choose\nn}
\dnu\dot\xmu\dmu\ynu, \nl
2.\ &\nn=\ss+\nu:\ &
-\sumnsp{-2} {\ss+\nu\choose\ss+\nu} {\ss+\mu+\nu\choose\ss}
\dot\xmu \d_{\nu+\mu}\ynu, \nle
3.\ &\ss=\nn+\mu:\ &
-\sumnsp{-2} {\nn+\mu+\nu\choose\nn} {\nn+\mu\choose\nn+\mu}
\d_{\mu+\nu}\dot\xmu\ynu, \nl
4.\ &\nn+\mu=\ss+\nu:\ &
-\sumrp{-2} {\rr+\mu+\nu\choose\rr+\nu} {\rr+\mu+\nu\choose\rr+\mu}
\dmu\dot\xmu \dnu\ynu,
\nonumber
\end{alignat}
where $\nn=\rr+\nu$, $\ss=\rr+\mu$ in the last line. 
When $\mu=\nu$, the conditions become $n_\nu\leq s_\nu+1$ and
$s_\mu\leq n_\mu+1$, together with $n_\rho=s_\rho$ for all 
$\rho\neq\mu=\nu$. The expression is now non-zero only in the first
three cases in the list above, whereas the fourth possibility is subsumed 
by the first one: if $\mu=\nu$, $\nn+\mu=\ss+\nu$ whenever
$\nn=\ss$.
The second and third cases cancel
the two first terms in (\ref{appLL}), which leaves us with
\bes
&& -\ktpi\int \Big\{ \sum_{\mu,\nu} \sumnp{-1} 
{\nn+\nu\choose\nn} {\nn+\mu\choose\nn} \dnu\dot\xmu\dmu\ynu +\nl
&&+ \sum_{\mu\neq\nu} \sumrp{-2} {\rr+\mu+\nu\choose\rr+\nu}
 {\rr+\mu+\nu\choose\rr+\mu} \dmu\dot\xmu\dnu\ynu \Big\} 
\label{ext1}\\
&=& -{k_4\/2\pi i}{N+p+1\choose N+1} \dnu\dot\xmu\dmu\ynu
- {k_4\/2\pi i}{N+p\choose N+1} \dmu\dot\xmu\dnu\ynu,
\eens
by Lemma \ref{LC}.

We now replace $\Lxi\mapsto\Lxi+T_{d\xi}$. That the regular terms
still yield a realization of $diff(N)$ is clear by Lemma \ref{Lemb},
so there remains to calculate the extension. Evidently,
\bes
\ext([\Lxi+T_{d\xi}, \Leta+T_{d\eta}]) &=& \ext([\Lxi,\Leta]) 
+ \ext([\Lxi, T_{d\eta}]) -\nl
&&-\ext([\Leta, T_{d\xi}]) + \ext([T_{d\xi}, T_{d\eta}]).
\eens
The first term is given by (\ref{ext0}) and (\ref{ext1}). The second
term follows from (\ref{xx1}):
\bes
\ext([\Lxi, T_{d\eta}]) &=&
-{k_3\dlt^\rho_\nu\/2\pi i} \Np{+1} \int \dmu\dot\xmu \drho\ynu 
\nlb{ext2}
&=& -{k_3\/2\pi i} \Np{+1} \int \dmu\dot\xmu \dnu\ynu.
\eens
The third term gives an identical contribution, whereas the fourth
contribution follows from (\ref{dXY}):
\be
\ext([T_{d\xi}, T_{d\eta}]) &=&
-{k_1\dlt^\rho_\nu\dlt^\si_\mu+k_2\dlt^\rho_\mu\dlt^\si_\nu\/2\pi i}
 \Np{} \int \drho\dot\xmu\dsi\ynu 
\label{ext3}\\
&=& -{k_1\/2\pi i}\Np{} \int \dnu\dot\xmu\dmu\ynu
-{k_2\/2\pi i}\Np{} \int \dmu\dot\xmu\dnu\ynu.
\eens
The total extension is thus the sum of (\ref{ext0}), (\ref{ext1}),
(\ref{ext2}) (twice), and (\ref{ext3}). This fixes the abelian charges 
$c_1$ and $c_2$ to the values given in Theorem \ref{T1}.

\subsection{$L\J$ bracket}
\bes
[L_f,\J_X] &=& \iint[ f(-\no{\dot\qmu\pmu} + F), 
 \summn{\mm\choose\nn} \d_{\mm-\nn}X_a J^{\mm a}_\nn] \nl
&=& \summn{\mm\choose\nn} \iint f \Big\{
 -\dot\qmu\d_{\mm-\nn+\mu}X_a J^{\mm a}_\nn \dlt
\nlb{appRJ}
&&+ \d_{\mm-\nn}X_a( J^{\mm a}_\nn \dot\dlt
 + {d_2\/4\pi i} \dlt^a \dlt^\mm_\nn (\ddot\dlt+i\dot\dlt) \Big\}\nl
&=& {d_2\/4\pi i} \dlt^a \summp{} 1 \int (\ddot f - i\dot f)X_a,
\eens
where we used that $\dot\qmu\d_{\mm-\nn+\mu}X_a = \d_{\mm-\nn}\dot X_a$.
The value of the abelian charge $c_6 = d_2\Np{}$ now follows from 
Lemma \ref{LA}.

\subsection{$L\L$ bracket}
Again we initially set $T_{d\xi}=0$:
\bes
[L_f,\Lxi] &=& \iint[ f(-\no{\dot\qrho p_\rho} + F),  
 \no{\xmu\pmu} - \xmu P_\mu
 + \summn {\mm\choose\nn}\d_{\mm-\nn}\xmu E^\mm_{\nn+\mu}] \nl
&=& \iint f \Big\{ 
 (-\no{\dot\qrho\drho\xmu\pmu} + \dot\qrho\drho\xmu P_\mu
- \summn {\mm\choose\nn}\dot\qrho\d_{\mm-\nn+\rho}\xmu E^\mm_{\nn+\mu})
\dlt\nl
&&+ \xmu\prho\dlt^\rho_\mu\dot\dlt - \xmu P_\mu\dot\dlt
+ \summn {\mm\choose\nn}\d_{\mm-\nn}\xmu E^\mm_{\nn+\mu}\dot\dlt \nl
&&+ \fpi(\dmu\xmu 
 +d_0 \summn {\mm\choose\nn}\d_{\mm-\nn}\xmu\dlt^\mm_{\nn+\mu})
 (\ddot\dlt+i\dot\dlt) \Big\}\nl
&=& 0 + \fpi(1+d_0\sumnp{-1}{\nn+\mu\choose\nn}) 
 \int (\ddot f - i\dot f) \dmu\xmu.
\eens
In the last step, we used that $\dot\qrho\drho\xmu = \dot\xmu$ to
eliminate all regular terms. The abelian charge $c_3$ has two 
contributions: one from the observer's trajectory and one from $F(t)$;
its value now follows from Lemma \ref{LB}.

The contribution from $T_{d\xi}$ follows from (\ref{appRJ}) by
replacing $d_2\dlt^aX_a \mapsto d_1\dmu\xmu$:
\be
[\Lxi, T_{d\xi}] 
= {d_1\/4\pi i} \Np{}\int (\ddot f - i\dot f)\dmu\xmu.
\ee

\subsection{$LL$ bracket}
Since $[F(s),{\dot\qmu(t)\pmu(t)}] = 0$, 
$L_f = \int (-f\no{\dot\qmu\pmu} + fF)$ consists of two commuting
Virasoro generators. The first consists of $N$ bosonic fields, each
contributing $+2$ to $c_4$, and the second contributes $c(N,p)$ by
assumption (\ref{Vir}). Hence $c_4 = 2N+c(N,p)$.

\section{Proof of Theorem \ref{T2}}
\label{appP2}
It is straightforward to show that the redefinitions (\ref{shift})
only affects the trivial cocycle in the $L\L$ and $L\J$ brackets.
Thus the unprimed operators satisfy the DGRO algebra (\ref{DGRO}), 
with the following modifications:
\bes
{[}L_f, \Lxi] &=& {c_3\/4\pi i} \int dt\ 
 \ddot f(t)\dmu\xmu(q(t)), \nle
{[}L_f,\J_X] &=& {c_6 \/4\pi i}\dlt^a 
 \int dt\ \ddot f(t) X_a(q(t)).
\eens

\subsection{$L\J$ bracket}
\bes
[L_f, \J_X] &=& \iint [f(-\no{\dot\qmu\pmu} + \dot\qmu P_\mu),
\summn {\mm\choose\nn}\dlt_{\mm-\nn}X_a J^{\mm a}_\nn] \nl
&=& \summn {\mm\choose\nn}  \iint f \Big\{
 -\dot\qmu \dlt_{\mm-\nn+\mu}X_a J^{\mm a}_\nn \dlt \nl
&&+ \dot\qmu \dlt_{\mm-\nn}X_a 
 ((J^{\mm-\mu a}_\nn - J^{\mm a}_{\nn+\mu})\dlt
 + {k^{a\emp}\/2\pi i}\dlt^\mm_{\nn+\mu}\dot\dlt ) \Big\} 
\label{appRJ2}\\
&=& \summn \Big\{-{\mm\choose\nn}+{\mm+\mu\choose\nn}
 -{\mm\choose\nn-\mu}\Big\} 
 \int f \dot\qmu \dlt_{\mm-\nn+\mu}X_a J^{\mm a}_\nn \nl
&&- {k^{a\emp}\/2\pi i}\sumnp{-1}{\nn+\mu\choose\nn}
 \int\dot f\dot\qmu\dmu X_a \nl
&=& 0+ {k_6\dlt^a\/2\pi i}\Np{+1}\int\ddot f X_a,
\eens
by Lemmas \ref{L2} and \ref{LA}. Hence $c_6 = 2k_6\Np{+1}$.

\subsection{$L\L$ bracket}
We first set $T_{d\xi}=0$.
\bes
[L_f, \Lxi] &=& \iint [f(-\no{\dot\qrho\prho} + \dot\qrho P_\rho),
\no{\xmu\pmu} - \xmu P_\mu 
 + \summn {\mm\choose\nn}\d_{\mm-\nn}\xmu E^\mm_{\nn+\mu}].
\eens
The regular terms yield
\bes
&&\int f\Big\{
-\dot\xmu\pmu + \dot\xmu P_\mu 
- \summn {\mm\choose\nn}\d_{\mm-\nn}\dot\xmu E^\mm_{\nn+\mu} +\nl
&&+ \summn {\mm\choose\nn}\dot\qrho \d_{\mm-\nn}\xmu 
( E^{\mm-\rho}_{\nn+\mu}-E^\mm_{\nn+\mu+\rho})\Big\} +\\
&&+\iint f\xmu (\pmu - P_\mu) \dot\dlt = 0,
\eens
by Lemma \ref{L2} and $\dot\qrho\drho\xmu = \dot\xmu$.

The extension has one contribution
$\fpi\int(\ddot f -i\dot f)\dmu\xmu$ from the observer's trajectory.
The rest of the extension becomes
\be
&&\iint f \dot\qrho\summn {\mm\choose\nn}\d_{\mm-\nn}\xmu 
\ktpi \dlt^\mm_{\nn+\mu+\rho} \dot\dlt \nl
&=& -{k_4\/2\pi i} \int \dot f \sumnp{-2} {\nn+\mu+\rho\choose\nn} 
 \dot\qrho\d_{\mu+\rho}\xmu \\
&=& {k_4\/2\pi i} \Np{+2} \int \ddot f \dmu\xmu,
\eens
where we used Lemma \ref{LE} in the last step.

The contribution from $T_{d\xi}$ follows from (\ref{appRJ2}) by
replacing $k_6\dlt^aX_a \mapsto k_3\dmu\xmu$:
\be
[\Lxi, T_{d\xi}] = {k_3\/2\pi i}\Np{+1}\int\ddot f\dmu\xmu.
\ee
Taking the three contributions together, we find that
\be
c_3 = 1 + 2k_4\Np{+2} + 2k_3\Np{+1}.
\ee

\subsection{$LL$ bracket}
We note that $\pmu(t) + P_\mu(t)$ satisfy the same Heisenberg algebra
as $\pmu(t)$:
\bes
[\pmu(s)-P_\mu(s), \qnu(t)] &=& \dlt^\nu_\mu\dlt(s-t), \nle
[\pmu(s)-P_\mu(s), \pnu(t)-P_\nu(t)] &=& 0.
\eens
Therefore, $L_f' = -\int f\no{\qmu(\pmu-P_\mu)}$ satisfies the same
algebraic relations as $L^0_f = -\int f\no{\dot\qmu\pmu}$ does, i.e. a
Virasoro algebra with central charge $c_4 = 2N$.

\end{document}